\documentstyle[emulateapj,psfig]{article}

\begin{document}

\newcommand{\be}{\begin{equation}}
\newcommand{\ba}{\begin{eqnarray}}
\newcommand{\ee}{\end{equation}}
\newcommand{\ea}{\end{eqnarray}}

\title {The Role of Heating and Enrichment in Galaxy Formation}   

\author{Evan Scannapieco\altaffilmark{1} \& Tom Broadhurst\altaffilmark{2}}
\altaffiltext{1}
{Department of Astronomy, University of California, Berkeley, CA  94720}
\altaffiltext{2}
{European Southern Observatory, Garching M\"{u}nich, Germany}

\begin{abstract}
  
We show that the winds identified with high-redshift low-mass galaxies
may strongly affect the formation of stars in more massive galaxies
that form later. With 3D realizations of a simple nonlinear growth model
we track gas shocking, metal enrichment, and cooling, together with
dark halo formation.  We show that outflows typically strip baryonic
material out of pre-virialized  intermediate mass halos, suppressing star
formation. More massive halos can trap the heated gas but collapse
later, leading to a broad bimodal redshift distribution, with a larger
characteristic mass and metallicity associated with the lower redshift
peak.  This scenario accounts for the observed bell-shaped luminosity
function of early-type galaxies, explains the small number of Milky-Way 
satellite galaxies relative to standard Cold Dark Matter proscriptions,
and provides a reasonable explanation for the lack of metal
poor stars in the solar neighborhood and the more general lack of
low-metallicity stars in other massive galaxies relative to ``closed-box''
models of chemical enrichment. Heating of the intergalactic medium by
early outflows should produce spectral distortions in the cosmic microwave
background that will be measurable with the next generation of
experiments.

 \end{abstract}



\section{Introduction}
 
It has long been recognized that the X-ray luminosity-temperature
($L_x - T$) relation of clusters does not obey the simple scaling laws
that would hold if clusters were formed from the collapse of unheated
primordial gas, and thus the gas within clusters is likely to have
been heated before the formation of the clusters themselves (Kaiser
1991).  Subsequent investigations have determined that preheating is
also necessary to explain the $L_x -T$ relation as a function of
cluster mass (Cavaliere, Menci, \& Tozzi 1999).

The high metallicity of cluster gas and the claimed over-abundance of
alpha elements (Gibson and Matteucci 1996; Lowenstein \& Mushotzky
1996) points to preheating by SNII driven winds (Renzini et al.\ 1993;
Trentham 1994; Nath \& Chiba 1995).  High
enrichment is not restricted to the most massive clusters, but appears
to be widespread, extending to groups of galaxies (Buote
2000). Empirically the epoch of this enrichment is now known to occur
at $z>0.5$, and approximately solar enriched
cluster gas has been detected
at redshifts as large as $z\sim1$ (Hattori et al. 1997).
Hence it is natural to suppose that pre-enrichment and preheating are
the consequence of very early and vigorous massive star-formation.

Cluster gas is evidently a sink for enriched and heated material, and
the most likely culprit for this enrichment is dwarf galaxies.
Theoretical work has shown that supernovae and OB winds in these
low-mass objects should lead to the production of energetic
outflows with temperatures on the order of $10^6$ K (Larson 1974;
Dekel \& Silk 1986; Vader 1986).  This behavior has been clearly
identified in studies of both local starbursting galaxies (Axon \&
Taylor 1978; Marlowe et al.\ 1995; Heckman 1997; Hunter et al.\ 1998;
Martin 1998) and spectroscopy of high-$z$ galaxies (
Franx et al.\ 1997; Pettini et al.\ 1998; 
Frye \& Broadhurst 1998; Warren et al.\ 1998). Whether these outflows lead
to a catastrophic loss of the interstellar gas however, is likely to 
depend on a number of factors (De Young \& Heckman 1994), and is a
subject of current investigations (Murakami \& Babul 1999; Mac Low \&
Ferrara 1999; Strickland \& Stevens 1999).

In the generally investigated hierarchical models of structure
formation such as the Cold Dark Matter model (CDM), the existence of
an era of widespread enrichment by outflows from dwarf galaxies is 
in fact quite natural, because low-mass galaxies are expected to form in
large numbers and at early times (e.g., White \& Frenk 1991).  Since
these early galaxies will be found  preferentially in the large-scale overdense
regions that later form clusters, dwarf outflows are the obvious
candidates for pre-cluster heating. 

The existence of large numbers of small galaxies at high redshifts is
also favored by observations, to help understand the steep number
counts and low luminosities of faint galaxies (Broadhurst, Ellis, \&
Glazebrook 1992), the sizes of faint galaxies in Hubble Deep Field
images (Bouwens, Broadhurst, \& Silk 1998a,b), and the small sizes of the
distant Lyman-break galaxies (Steidel et al.\ 1999).  Locally,
however, the space density of dwarf galaxies relative to
massive galaxies is far less than predicted on the basis of the steep
Press-Schechter slope for the faint end of the mass function (Ferguson
\& Binggeli 1994), prompting theoretical studies of the disruption of
dwarves by tidal forces from neighboring objects (Moore et al. 1998),
external UV radiation (Kepner, Babul, \& Spergel 1997; Norman \&
Spaans 1997; Corbelli, Galli, \& Palla 1997), and  catastrophic mass
loss during outflows (Larson 1974; Dekel \& Silk 1986; Vader 1986).

While many of these mechanisms may have had an impact on the formation
of dwarf galaxies, relatively little attention has been directed
towards the influence of outflows on neighboring galaxies.  As the
earliest galaxies to form were highly clustered (Kaiser 1984) and
typical outflow temperatures and velocities were much larger than the
virial temperatures and velocities of these galaxies, it is likely
that dwarf galaxies were strongly influenced by their neighbors
(Scannapieco, Ferrara, \& Broadhurst 2000).  Similarly, the sources
responsible for pre-enrichment of the intracluster medium may well
have enriched larger protogalaxies, with important consequences for
their metallicity histories and cooling times.

In this work, we conduct an idealized investigation of the impact of
dwarf outflows on the history of galaxy formation.  While the
consequences of homogeneous heating on galaxy formation have been
examined in the past (Blanchard, Valls-Gabaud, \& Mamon 1992), the
inhomogeneous nature of this process and the associated pre-enrichment
of galaxies have not been addressed.  Our aim is not, however, to
construct a complete model of galaxy formation, and thus we do
not track processes such as the formation of second-generation stars
in galaxies, production of dust, transfer of angular momentum, or the
structure of the interstellar medium.  Rather we focus on
the properties of the inhomogeneously heated and enriched
intergalactic medium (IGM) out of which galaxies coalesced, the likely
consequences of galaxy formation in this environment, and to what
degree these issues must be accounted for 
within more detailed simulations of galaxy formation.

The structure of this work is as follows.  In \S2 we review the
observational evidence that leads us to consider a model in which
widespread dwarf outflows shocked and enriched the medium out of which
larger galaxies formed.  In \S3 we describe a simple numerical Monte
Carlo code that we use to asses the overall features of such a model.
The results of our simulations are given in \S4 in which we examine
which aspects of galaxy formation are most sensitive to the presence of
outflows. In \S5 we discuss the limitations of our modeling,
and conclusions are listed in \S6.

\section{Observational Evidence}

In this section we review the observational evidence that points 
towards widespread early enrichment by dwarf galaxies.   We show 
that the chemical and thermal properties of nearby galaxy clusters 
are both suggestive of such pre-enrichment, as is the weak evolution
of these properties with redshift.  We show that
such a picture is consistent
with our knowledge of the intergalactic medium as well, and helps
to explain the relatively high metallicity seen in many Ly
absorption systems.  The presence of outflows has been identified
in emission line  studies of high-$z$ galaxies, and several nearby
dwarf galaxies have been caught ``red-handed'', surrounded by 
clouds of enriched gas with temperatures greatly exceeding 
virial.  Finally, we review some unsolved questions in 
galaxy formation that may depend on preheating and enrichment
and serve to motivate our exploratory numerical studies.

\subsection{Galactic Outflows and  the Intra Cluster Medium }

Many cluster properties can be understood in the context of
self-similar models (Kaiser 1986, 1990) which are exact for power-law
initial fluctuation spectra and a reasonable approximation for more
realistic spectra such as that of the Cold Dark Matter (CDM) model.
While the optical properties of galaxies in clusters are consistent
with these predictions, the slope of the X-ray luminosity-temperature
relationship of the hot intra-cluster gas is too steep and slowly
evolving to be understood in this context.

A convincing explanation of $L_x-T$ evolution has been made by Kaiser
(1991) who speculated that at an epoch before the formation of
present-day clusters, the gas which formed the intracluster medium
(ICM) was preheated, injecting sufficient energy to
expel gas from the small potential wells that were non-linear at early
times.  By assuming that this heated gas later cooled adiabatically
onto the large cluster-size potential wells, he was able to simply
reproduce the observed X-ray properties of clusters.  Numerous
investigations of preheating and $L_x-T$ evolution have also reached
similar conclusions (David, Jones, \& Forman, 1996; Mushotzky \&
Scharf 1997; Eke, Navarro, Frenk 1998) as have studies of the $L_x-T$
relation in groups of galaxies (Cavaliere, Menci, \& Tozzi 1999).

The origin of this preheating is most naturally due to Supernovae
Type-II (SNeII) activity in the dwarf starburst population.  Several
studies of expanding HI gas in nearby dwarf galaxies show clear
evidence of dense expanding shells with velocities above
$15$ km/s (Marlowe et al.\ 1995; Heckman 1997;
Hunter et al.\ 1998; Martin 1998). ${\rm Ly}\alpha$ studies
confirm that similar outflows are present in higher $z$ galaxies at
redshifts of order $\sim 3$ (Pettini et al.\ 1998) and even higher
(Frye \& Broadhurst 1998; Warren et al.\ 1998
), where the overall space density
of small starbursting galaxies is larger.

A natural prediction of such a picture would not only be the heating
of the ICM by the dwarf population, but the enrichment of this gas by
metals expelled by the SNeII powering the galactic winds.  In fact,
the metallicity of the intracluster medium is observed to be quite
high and constant ($\approx 0.3 Z_\odot$) over a large range of
cluster masses (Renzini 1997).  This roughly constant value is
evidence that the clusters have neither lost or gained a large amount
of baryons during their evolution, as massive cluster outflows or
accretion of pristine gas would lead to a large scatter.  Furthermore,
this metallicity is roughly the same today as observed at $z \approx
0.3$ and perhaps even at $z \approx 1 $ (Hattori et al.\ 1997),
suggesting that IGM enrichment took place early on in the
lifetime of the clusters (Renzini 1999).

Several authors have investigated the possibility that this enrichment
is due to the present cluster members and that most of the X-ray
emitting hot gas in clusters is due to outflows from galaxies that
populate them today.  Okazaki et al.\ (1993) considered gas ejection
from bright, elliptical galaxies in clusters, concluding that these
galaxies could contribute no more than 10\% of the ICM.  Trentham
(1994) suggested that the ICM instead may be primarily due outflows
from precursors to the observed populations of dwarf galaxies in
clusters, but Nath and Chiba (1995) concluded that the metallicities
generated in this type of scenario were sufficient to explain only
clusters with low-metallicity gas.

Perhaps most convincingly Gibson and Matteucci (1996) have studied
the possibility that the elliptical galaxies observed
in clusters today were responsible for the majority of ICM enrichment.  By
developing models of galactic winds consistent with the observed
properties of cluster ellipticals they concluded that even in their
``maximal models'' in which all the gas returned by dying stars is
ejected into the ICM, neither the giant elliptical, the dwarf
spheroidals, or both these populations combined can be responsible for
more that 33/38\% of the ICM.  And yet this gas is hot and metal
rich.

A final piece of evidence as to the origin of ICM metals comes from
studies of relative abundances.  While for many years, studies of
metals in clusters were limited to measurements of iron abundances,
the X-ray {\it Advanced Satellite for Cosmology and Astrophysics}
(ASCA) (Tanaka, Inoue,\& Holt 1994) provided the first opportunity to
study other metals.  This allowed Loewenstein \& Mushotzky (1996) to
observe excesses of alpha-elements such as O, Mg, and Si, which
indicate that most of the ICM enrichment was due to SNeII rather than
Type Ia (SNeIa).  This result was later contested by Ishimaru \&
Arimoto (1997), who claimed that SNeIa enrichment could be responsible
for 50\% or more of ICM metals, based on the solar `meteoritic'
metallicity rather than the solar coronal gas.  Gibson, Loewenstein, 
\& Mushotzky (1997) pointed out, however,
 that this result was specifically linked to the
SNeII yield used in their analysis.  While they were unable
to rule out Ishimaru \& Arimoto's claim within theoretical
uncertainties, they showed that more recent SN models that treat
convection and mass-loss, reduce the fractional contribution ICM by
SNeIa to less than 5\%.   Renzini (1997) on the other hand
while confirming the importance of early SneII, has argued
that later enrichment by significant numbers of SNeIa are
required to produce the observed high level of Fe relative to alpha
elements.

In summary then, metallicities studies of clusters indicate that the ICM
was enriched early on, that this enrichment is too high to be due to
purely to the observed cluster members, and that the relative 
abundances of metals are suggestive of widespread SNeII enrichment; 
all indicative of enrichment by high-redshift dwarf galaxies.

\subsection{Galactic Outflows and the Intergalactic Medium}

The first galaxies to form would have necessarily had a huge impact
not only in regions that would later form clusters, but on the
Intergalactic Medium (IGM) as a whole.  In standard schemes for
hierarchical structure formation, the first baryonic objects formed at
redshifts $\sim40$ and with masses of order $10^5 M_\odot$, somewhat
smaller than the dwarf galaxies observed today (Haiman, Thoul, \& Loeb
1996).  While several authors have shown that the total flux of UV
photons from stars within these objects is sufficient to reionize the
universe (Couchman \& Rees 1986; Fukugita \& Kawakasi 1994; Shapiro,
Giroux, \& Babul 1994), these objects are likely to suppress their own
formation long before this occurs.  As molecular hydrogen is
easily photo-dissociated by 11.2-13.6 eV photons, to which the
universe is otherwise transparent, the emission from the first stars
quickly destroys all avenues for cooling by molecular line emission.
This quickly raises the minimum virial temperature necessary to cool
effectively to approximately 10,000 K, suppressing the further
formation of objects with masses $\lesssim 10^8 M_\odot$ (Haiman Rees,
\& Loeb 1997; Ciardi et al.\ 1999).

Thus reionization, if achieved by galaxies, was relatively late ($z
\lesssim 15$) and associated with objects of similar size or larger
than dwarf galaxies.  This high mass scale, implies that the IGM phase
transition must have occurred in a clumpy and inhomogeneous manner,
with important implications for small-scale microwave background
anisotropies (see eg.\ Aghanim et al.\ 1996; Miralda-Escude, Haehnelt, \& Rees
 2000; Scannapieco 2000).

Early enrichment and heating is similarly expected to be
inhomogeneous, following the spatial distribution of the first
galaxies which are restricted to rare overdense regions. Hence
measurements of a low average metallicity ($\sim 0.01 Z_\odot$) of the
Ly-forest at high redshift (Songalia \& Cowie 1996; Savaglio 1997;
Songalia 1997) may be viewed not so much as evidence for a low
metallicity intergalactic medium, as an inhomogeneous one.
High-redshift regions of higher density that later develop into
clusters would therefore represent the most polluted volumes, where one
would naturally expect a significantly higher mean metallicity.

Direct evidence that the IGM is enriched by galaxy outflows
may be inferred from detailed observations of 
local dwarf spheroidal galaxies.  For example, 
ASCA observations (della Ceca et
al.\ 1996) show that the star-forming dwarf galaxy NGC 1569 is
surrounded by a hot ($8 \times 10^6$ K) halo of gas whose temperature
greatly exceeds the virial temperature of the galaxy and whose line
strengths are consistent with $0.25 Z_\odot$.  Similarly, X-ray
observations of the dwarf irregular NGC 4449 by Bomans, Chu, \& Hopp
(1997) show that it is embedded in a supergiant shell of $\sim 2
\times 10^6 $K gas with a metallicity of $\sim 0.3 Z_\odot.$

\subsection{Galactic Outflows and Galaxy Formation}

\begin{figure}
\centerline{ 
\psfig{file=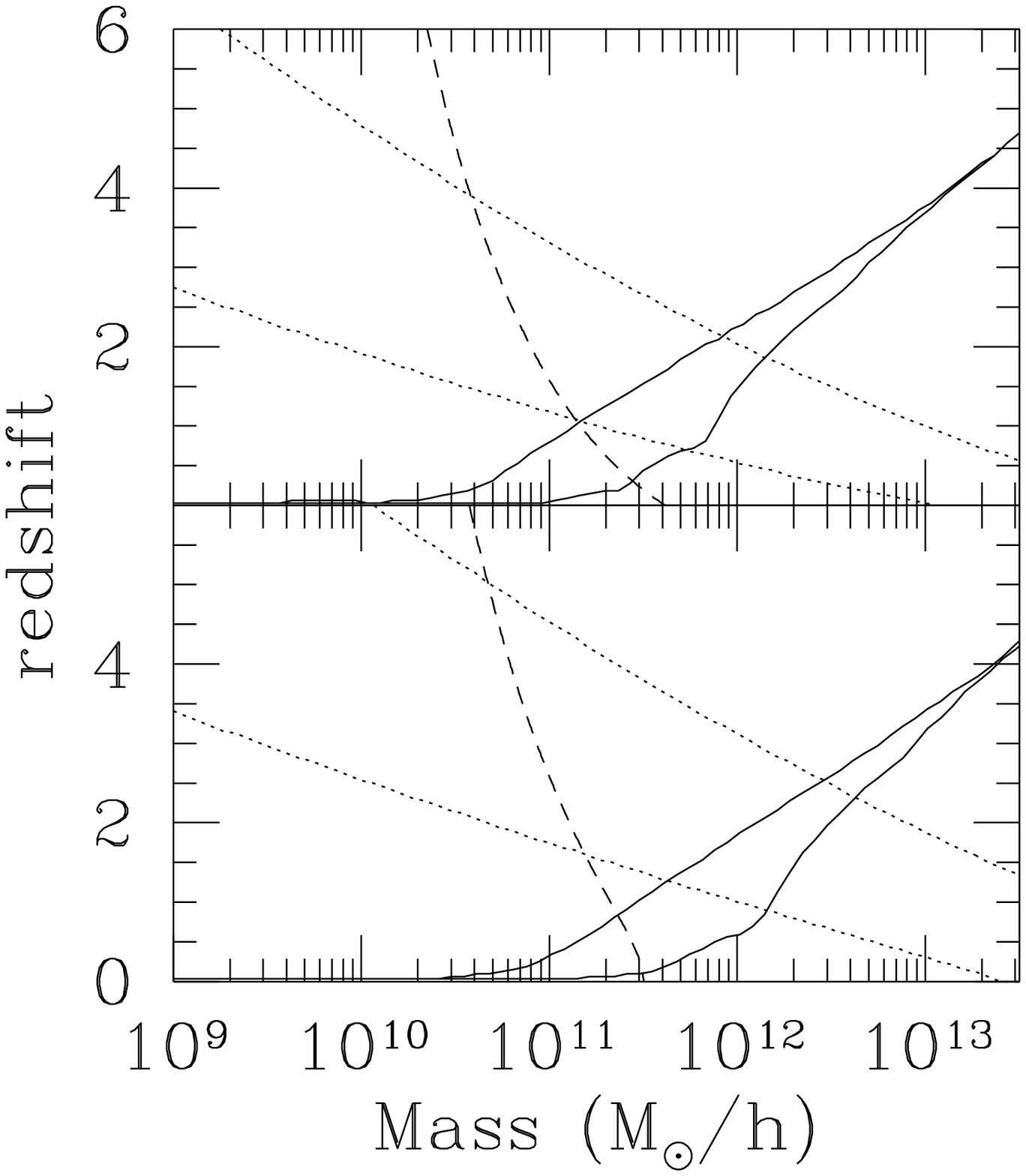,width=3.8in}}
\caption{The impact of galactic outflows on galaxy formation.  
The solid lines show the lowest $z$ at which a halo can collapse and
still form a galaxy by $z = 0$, illustrating the enhanced
cooling of metal-enriched virialized objects versus objects formed
of primordial gas.   In each pair the upper line
corresponds to primordial gas, and the lower line to gas that has been
enriched to $0.1 \, Z_\odot$.  The dotted lines show the collapse
redshift of the $2\sigma$ (upper line) and $1\sigma$ peaks (lower
line), indicating the redshift at which objects of various mass
scales form.
The dashed line corresponds to a fixed virial temperature of
$5 \times 10^5$ K, typical of outflows, and below which
outflows are capable of suppressing galaxy formation.
The upper panel lines are
calculated in a $\Omega_0 = 1$ CDM universe 
 ($\Omega_0 = 1$, $\Omega_\Lambda = 0$,
$\Omega_b = 0.07$, $\sigma_8 = 0.6$, and $\Gamma = 0.44$)
and the lower panel lines are for a 
$\Lambda$CDM model ($\Omega_0 = 0.35$, $\Omega_\Lambda = 0.65$,
$\Omega_b = 0.06$, $\sigma_8 = 0.87$ and $\Gamma = 0.18$).}
\label{fig:cc2}
\end{figure}

The presence of an inhomogeneous IGM with strong temperature and 
metallicity fluctuations creates quite a different environment for
galaxy formation than the primordial conditions often assumed.
This becomes clear when one contrasts the virialization and
cooling of primordial gas to that of gas pre-enriched and heated
by dwarf outflows.
In Figure \ref{fig:cc2} we replot a classic comparison first
made in Dekel and Silk (1986).
Collapse redshifts of $1\sigma$ to $2\sigma$ dark matter halos 
(dotted lines) are calculated from 
linear theory for both a flat CDM ($\Omega_0 = 1$, $\Omega_\Lambda = 0$,
$\Omega_b = 0.07$, $\sigma_8 = 0.6$, and $\Gamma = 0.44$) and a
$\Lambda$CDM model ($\Omega_0 = 0.35$, $\Omega_\Lambda = 0.65$,
$\Omega_b = 0.06$, $\sigma_8 = 0.87$ and $\Gamma = 0.18$), assuming
spherical collapse.  
The solid lines show the lowest redshift 
that a sphere of gas can virialize and still have time to
cool and form a galaxy by $z = 0$, as calculated by the cooling models
described in \S3.2.  In each pair the upper line corresponds to
primordial gas, and the lower line to gas that has been enriched to
$0.1 Z_\odot$.  Here we see that even modest enrichment by outflows
has the potential to greatly accelerate galaxy formation on the
$\sim 10^{12} M_\odot/h$ scale due to the long cooling times
of large clouds without the additional avenue for cooling
afforded by line emission by metals.


Also on this plot, we show the mass of halos associated with a virial
temperature of $5 \times 10^5$K, typical of galaxy outflows (dashed
line).  This serves as an estimate of the smallest mass of galaxies
that can form in areas impacted by outflows without being disrupted by
shocks. This suggests that while high-mass galaxy formation may be
enhanced by dwarf outflows, the formation of dwarf galaxies has the
potential to be suppressed.

The features shown on this plot invite comparison with the properties
observed in elliptical galaxies, which are preferentially found in the
most enriched regions in the universe, and whose mass functions are
biased to large values relative to the rest of the galaxy population.
It is also interesting to note that various analytical 
and N-body studies of cold dark matter models (Kauffmann,
White, \& Guiderdoni 1993; Klypin et al.\ 1999; Moore et al.\ 1999)
have shown that $\sim 50$ satellites with circular velocities
$\sim 20$ km/s should be found within 600 kpc of the Galaxy,
while only 11 are observed.  
Again, this lack of local dwarf galaxies
is suggestive of galaxy formation in an intergalactic
medium that has been impacted by outflows.

Closer to home, the abundance distribution of long-lived stars in
the solar neighborhood shows far too few low-metallicity stars compared
with a simple ``closed box'' model of galactic chemical evolution.  This is
the long standing G-dwarf problem, first pointed out by van de Bergh
(1962) and Schmidt (1963).  The sudden drop in the number of G-dwarfs
with metallicities below $\sim 0.1 Z_\odot$ has lead to a number of
models in which an initial production spike or ``prompt initial
enrichment'' adds metals to the gas out of which the majority of stars
form (e.g., Truran \& Cameron 1971; Ostriker \& Thuan 1975; K\"oppen
\& Arimoto 1990). 
This {\em ad hoc} floor to the metallicity then allows
good fits to the Galactic stellar data. 

Similar pre-enrichment may also be necessary to explain the
metallicity distribution of stars outside our own galaxy.  Integrated
spectra of elliptical galaxies have been studied with care (Worthey,
Dorman, \& Jones 1996) concluding rather
puzzlingly that such galaxies do not have a closed-box history but
again show the need for a ``floor'' of around $0. 1Z_\odot$ 
(Thomas, Greggio, \& Bender 1999). 

\section{Cosmological Monte Carlo}

The great preponderance of observational clues and pointers suggests
that we should seriously consider a scenario in which the formation of
modern-day galaxies was preceded by an era of IGM enrichment by a
high-redshift population of dwarf starbursting galaxies.  Note that
from a theoretical point a view, such a model is not {\em ad hoc}, but
rather a consequence of hierarchal structure formation.  As
smaller objects form early and are able to suppresses further
small-scale formation, it is only natural that larger galaxies would
have formed in a second wave of collapse within this pre-enriched
medium.

The density dependence of the effect of outflows on galaxy formation
requires a 3-D calculation. To date only the average properties of the
IGM as a function of time have been explored in one dimensional
calculations by Blanchard, Valls-Gabaud, \& Mamon (1992) and by Nath
and Trentham (1997). Such averages are appropriate to the IGM as a
whole, but underestimate the effect on galaxy formation. 
Thus for example, the outflow model proposed by Nath and Trentham (1997) 
produces metallicities $\sim 
.01 Z_\odot$, while the observed metallicities of cluster
gas and elliptical galaxies are comparable to solar abundance (Renzini
1993).
 
In order to determine the impact of outflows on galaxy formation,
we have developed a simple cosmological Monte Carlo code, in which we
realize a small volume of the universe and study the linear evolution
of objects in the mass range  
$2 \times 10^8 \Omega_0/h {\rm M}_\odot \leq M \leq 
1 \times 10^{13} \Omega_0/h {\rm M}_\odot$ 
within it, with reasonable extensions to 
deal with collapse.
  Our philosophy is not to pretend that we can possibly reproduce in
detail the complicated processes of dark matter halo collapse, galaxy
formation, and gas infall and expulsion, but rather to explore a model
that captures the essential features and predicts with some confidence
rough magnitudes and trends that can be compared with observations.

\subsection{Collapse of Dark Matter Halos}

We choose for our simulation a cubic comoving volume of $(12{\rm
Mpc}/h^{-1})^3$ divided into $512^3 $ cells and with periodic boundary
conditions.  On this mesh we construct a cosmological linear
overdensity field $\delta({\bf x},z) = \rho({\bf x},z)/\ {\bar
\rho}(z)$ where ${\bar \rho}(z)$ is global average density of the
universe, ${\bf x}$ is a position in comoving coordinates, and $z$ is
the redshift.  Transforming into Fourier space the density field can
be reexpressed as
$   \tilde \delta({\bf k},z) \equiv \int \,d^3 {\bf x} \exp(-i
     {\bf k} \cdot {\bf x})\, \delta({\bf x},z)$.
From linear theory, the evolution of the density field as a function
of time is given simply by $\tilde \delta({\bf k},z) = \tilde
\delta_0({\bf k}) D(z)/D_0$ where $D(z)$ is the dimensionless growth
factor, $\tilde \delta_0({\bf k}) \equiv \tilde \delta({\bf k},0)$,
and $D_0 \equiv D(0).$

In the CDM model, the Gaussian random field $\tilde \delta({\bf k})$
can by constructed by randomly choosing modes such that the variance
is given by
$
\langle \tilde{\delta}({\bf k}) \tilde{\delta}({\bf k'}) \rangle =
\langle \tilde{\delta}({\bf k}) \tilde{\delta}^*(-{\bf k'}) \rangle =
(2 \pi)^3 \tilde{\delta}({\bf k} + {\bf k'}) P({\bf k})$.
The power spectrum is then 
\be
     P(k) = {2 \pi^2} \delta_H^2 k^n T^2(k_p\,{\rm
     Mpc}/h \Gamma),
\label{eq:powerspectrum}
\ee
where $T(q)$ is the CDM transfer function given by Eq.\ (G3) of
Bardeen et al.\ (1996), and we choose $a_0 H_0=c$, such that $k_p= k/
a_0 =k H_0/c$ is the physical wave number.  The shape parameter
$\Gamma$ is defined as $\Gamma \equiv \Omega_0 h \exp(-\Omega_b -
\Omega_b/\Omega_0)$ (Sugiyama 1995) and is observed be to $\Gamma =
0.23^{+0.042}_{-0.034}$ given a spectral index of $n = 1$ (Vianna \&
Liddle 1996).  We fix the normalization factor, $\delta_H$,
against the number abundance of clusters.  
The variance of the mass inclosed in a sphere of radius $R$ is
\be
\sigma^2(R) = \frac{1}{2 \pi^2} \int_0^\infty k^2 dk P(k) W^2(kR).
\label{eq:sig}
\ee
If we choose a spherical top-hat window function defined by
$W(x) \equiv 3 \left[ \frac{\sin(x)}{x^3} - \frac{\cos(x)}{x^2}\right]$ 
then we can normalize our fluctuation spectrum by setting
$\sigma_8 \equiv \sigma (8 \, {\rm Mpc} \, h^{-1}) = 
(0.6 \pm 0.1) \Omega_0^{-C(\Omega_0)},$
where $C(\Omega_0) = .36 + 0.31\Omega_0-0.23 \Omega_0^2$ in the open
case and $C(\Omega_0) = .59 + 0.16\Omega_0 -0.06 \Omega_0^2$ if we take
($\Omega_0+ \Omega_\Lambda = 1$) (Vianna \& Liddle 1996).  

Having constructed $\tilde \delta_0({\bf k})$ in this manner, we then convolve
it with window functions with lengths scales corresponding
to ten masses arranged logarithmically from 
$M_{1}  \equiv 2 \times 10^8 \Omega_0/h M_\odot$ to 
$M_{10} \equiv  10^{13} \Omega_0/h M_\odot$, to obtain
$ \delta_0^{M_1}({\bf x}) \equiv
\delta_0({\bf x},R_{2 \times 10^8 \Omega_0/h M_\odot}) = 
  \int \,{d^3 {\bf k} \over (2\pi)^3} \,
\exp(i{\bf k} \cdot {\bf x})\, \tilde \delta({\bf k}) W(R_{2 \times 10^8 
\Omega_0/h M_\odot} k)$,
$\delta_0^{M_2}({\bf x}) \equiv \delta_0({\bf x},R_{6.6 \times 10^8 \Omega_0/h M_\odot})$, 
... $\delta_0^{M_{10}}({\bf x})$, the {\it linear} density field at $z = 0$ 
smoothed at each mass scale.  Here
the minimum mass of our simulations is set by molecular cooling
constraints discussed further below.

From the spherical collapse model we can identify the linear
overdensity $\delta_{\rm sc}(z)$ at which the true density field has
virialized.  For flat models this value is $\simeq 1.69$ at all times
while in open cases it is a weakly decreasing function of redshift as
fitted in Appendix A of Kitayama
\& Suto (1996).  For each peak in the linear overdensity field
such that $\delta^{M_i}_0({\bf x},R) > \delta_{\rm sc}(z=0)$, we identify a collapse
redshift $z_{\rm sc}$ such that
\be
\delta^{M_i}_0 ({\bf x}) = \delta_c(z_{\rm sc})
\frac{D_0}{D(z_{\rm sc})}.
\ee
This approach has the limitation that it becomes inaccurate 
as $\sigma(R) D(z)/D_0$ approaches $\delta_{\rm sc}$. Thus while the
spherical collapse model works well at determining the number and 
spatial clustering of rare objects (Lacy \& Cole 1993; 
Mo, Jing, \& White 1996, 1997) it fails to agree with numerical simulations
for more common low-mass halos (Lacey \& Cole 1994; 
Sheth \& Tormen 1999).  Furthermore, the object-by-object identification
of linear peaks with collapsed objects has been shown to be 
unreliable, even in cases in which the statistical properties are in
good agreement with N-body simulations (Bond et al.\ 1991).

Sheth, Mo, \& Tormen (1999) have shown that
these predictions can be improved by accounting for the ellipticity
of collapsing clouds caused by tidal forces from nearby collapsing peaks.
By including a correction factor for the critical linear overdensity, 
they have been able not only to improve statistical predictions over the
spherical model, but do well on an object-by-object basis.  As our 
Monte-Carlo approach depends on this object-by-object identification,
we can greatly improve the accuracy of our simulations by comparing
each peak with $\delta_{\rm ec}(z)$, such that
\ba
\delta_0^{M_i} = & {\delta_{\rm ec}(z_{\rm ec})} 
\frac{D_0}{D(z_{\rm ec})}  \qquad \qquad \qquad \qquad \qquad \qquad 
\nonumber \\
 = & {\delta_{\rm sc}(z_{\rm ec})} 
\frac{D_0}{D(z_{\rm ec})} 
\left[ 1 + \beta 
\left( \frac{\sigma^2(R)}{\delta^2_{\rm sc}(z_{\rm ec})}
\frac{D(z_{\rm ec})^2}{D^2_0} \right)^\gamma
\right],
\label{eq:ellcriteria}
\ea
where $\beta = 0.47$ and $\gamma = 0.615$.
 
Having identified the collapsed peaks by either algorithm, we then
arrange them in order of decreasing collapse redshift,
$z_{\rm c} = z_{\rm sc}$  or $z_{\rm ec}$, to obtain a list of 
candidate points and redshifts for collapsed halos at each of the
different mass scales:
${\cal X}^{M_i}_{\rm candidate} 
\equiv \{ {\bf x}_1^{M_i}, {\bf x}_2^{M_i},...  \}
\,\,\, {\rm and} \,\,\, 
{\cal Z}^{M_i}_{{\rm c, candidate}} \equiv 
\{z_{{\rm c},1}^{M_i}, z_{{\rm c},2}^{M_i},... \}.$
From this list we exclude unphysical points corresponding to all halos
that collapse within a previously collapsed halo at the same or
greater mass scale.  That is we remove all points such that
\be
M_i \leq M_j, \,\,\,  ||{\bf x}_k^{M_i}-{\bf x}_l^{M_j}|| < R_{M_j}, 
\,\, {\rm and}  \,\, z^{M_i}_{{\rm c},k} < z^{M_j}_{{\rm c},l},
\ee
where the distance 
is calculated accounting for the periodicity of the simulation volume.

With these points excluded, we then have the full history of the
collapsing dark matter halos at each of the various mass scales:
${\cal X}^{M_i} \equiv \{{\bf x}_1^{M_i}, {\bf x}_2^{M_i},...  \}$, and 
${\cal Z}^{M_i}_{\rm c} \equiv \{z_{{\rm c},1}^{M_i}, z_{{\rm c},2}^{M_i},... \}$.
The merger history of these objects is also simply calculated
by searching for all points such that
\be
M_i < M_j, \,\,\,  ||{\bf x}_k^{M_i}-{\bf x}_l^{M_j}|| < R_{M_j}, 
\,\, {\rm and} \,\, z_{{\rm c},k}^{M_i} > z_{{\rm c},l}^{M_j}.
\ee
Each such case indicates that the halo centered at ${\bf x}_k^{M_i}$
was absorbed into the larger halo centered at ${\bf x}_l^{M_j}$
at a redshift of $z_{{\rm c},l}^{M_j}.$

\subsection{Comparison with Analytical Results}

At each redshift, the number density of collapsed halos in the 
simulation can be compared to analytical predictions.  
For the spherical collapse model this gives
(Press \& Schechter 1974) 
\ba
\frac{d n_{\rm sc}(M,z)}{dM} & = & -\sqrt{\frac{2}{\pi}} \frac{\rho(z)}{M}
\frac{\delta_{\rm sc}(z) D_0}{\sigma(M)^2 D(z)} \nonumber \\
& & \exp \left( 
-\frac{\delta^2_{\rm sc}(z) D_0^2}{2 \sigma^2(M) D(z)^2}
\right) 
\frac{d \sigma(M)}{dM}.
\label{eq:press}
\ea
Strictly speaking, Eq.\ (\ref{eq:press}) corresponds to the number of peaks
as determined by a sharp $k$-space filter, rather than the number
determined by the top-hot filter used in our code.  Equating these 
quantities has become somewhat of a common practice however, and
has been shown to be a good estimate.

\begin{figure}
\centerline{ 
\psfig{file=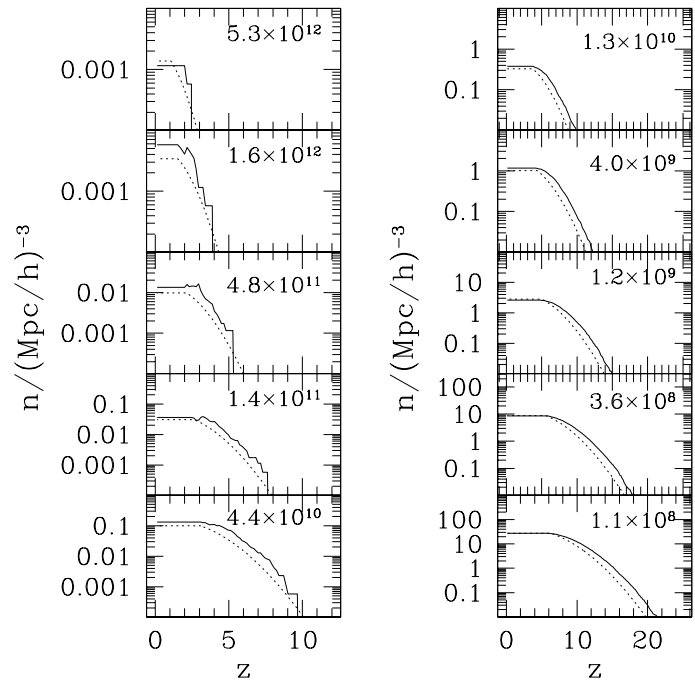,width=3.9in}}
\caption{ 
Check of our Monte Carlo space densities.  The
solid lines correspond to the 3-D Monte Carlo space density of halos at
each redshift and the dotted lines correspond to the analytical
Press-Schechter number density.  The panels are labeled by mass in
units of $M_\odot$.  Here we see that over the reliable range of
redshifts our method is in reasonable agreement with analytical
predictions on all scales.}
\label{fig:counts}
\end{figure}

In Figure \ref{fig:counts} we compare the number density of objects
in our simulation as function of redshift with the number  predicted
by Eq.\ (\ref{eq:press}).  
As both the Press-Schechter 
integral, and our peak finding algorithm become inaccurate at late times, we
exclude all peaks that  collapse with a redshift lower that a minimum
value, $z_{\rm min}(M_i)$ such that 
$\sigma(M_i) D(z_{\rm min}(M_i))/D_0 = 1.3 \delta_{\rm sc}.$ 
We consider the $\Lambda$CDM model
shown in Figure \ref{fig:cc2},  ($\Omega_0 = 0.35$, $\Omega_\Lambda = 0.65$, 
$\Omega_b = 0.06$, $\sigma_8 = 0.87$ and $\Gamma = 0.18$, $h=0.65$).


Here we see that the number of identified objects agrees well at most
mass scales and redshifts.  At the smaller scales, there is some
discrepancy in the collapse time of objects, with the code slightly
over-predicting the number of objects formed at very early times.  
At larger-range scales, the Monte-Carlo approach does
the best, matching the Press-Schechter predictions at all
redshifts.  Finally, at the highest mass scales, the low number of
objects introduces significant statistical noise in our comparison.

The spatial distribution of collapsed halos can also be
compared to analytical predictions.  If we define the correlation
function at a smoothing scale $R$ as the Fourier transform of the
power spectrum smoothed by an appropriate top-hat window function,
$\xi_R(r) \equiv \frac{1}{2 \pi^2} \int_0^\infty k^2 dk P(k) W^2(kR)
\frac{\sin(k r)}{k r}$,
then the correlation of collapsed peaks of scale $R$ can be 
approximated by (Kaiser 1984)
\ba
1 + \xi_{R,\nu}(r) =& \left(\frac{2}{\pi}\right)^{1/2} \,
\left[ {\rm erfc}(\nu/2^{1/2}) \right]^{-2} \, \times 
\qquad \qquad \nonumber \\
 & \int_\nu^\infty dy e^{-y^2/2} {\rm erfc}
\left[ \frac{ \nu - y \xi_R(r)/\xi_R(0)}
{\sqrt{2 - 2\xi_R^2(r)/\xi_R^2(0)}} \right],
\label{eq:kaiser}
\ea
where erfc is the complimentary error function and
$\nu \equiv \delta_c D_0/\sigma(R) D(z).$  Note that this expression
arises from considering the fraction of Gaussian-distributed 
objects above a certain threshold, rather than from the more sophisticated
excursion set approach (Bond, Efstathiou, \& Kaiser 1991; Lacy \& Cole
1993) used to derive Eq. (\ref{eq:press}).  

\begin{figure}
\centerline{ 
\psfig{file=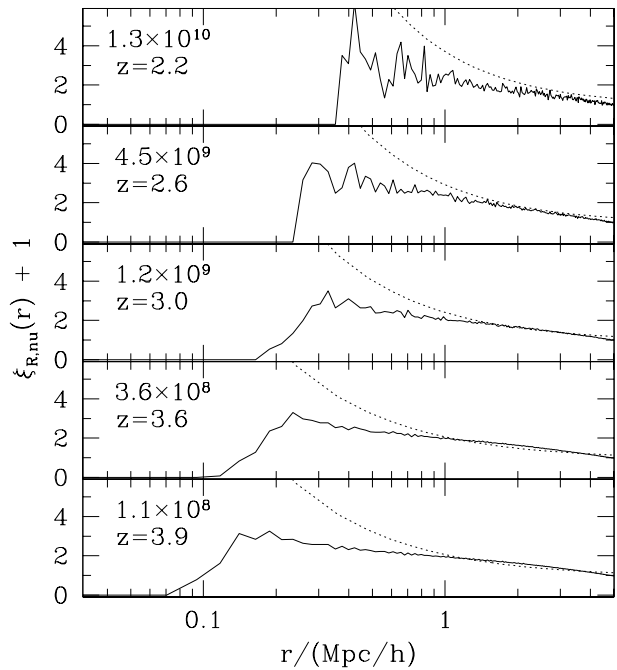,width=3.8in}}
\caption{
Comparison of analytical (solid lines) and Monte-Carlo (dotted lines) 
correlation functions at various mass scales labeled in units of
$M_/odot$.  The objects are calculated from the spherical 
collapse model and at redshifts such that 
$\nu \equiv \delta_c D_0/\sigma(R) D(z) = 1.5$.
While both the analytical and numerical curves must be  different
at small values of $r$ because of the manner in which they are 
calculated, good agreement exists 
at intermediate distances for all mass scales.}
\label{fig:eta}
\end{figure}

In Figure \ref{fig:eta} we compare the correlation function as given in
Eq.\ (\ref{eq:kaiser}) to that derived from the distances between the
collapsed peaks at the five lowest mass-scales in our simulation,
with redshifts chosen such that $\nu = 1.5$ at each length scale.
Eq.\ (\ref{eq:kaiser}) is intended to represent the correlations between
collapsed objects at or above a particular mass scale corresponding
to the smoothing length scale.  In practice, however, it makes little 
difference whether we compute $\xi_{R_{M_i}}(r)$ considering distances
between all objects with masses above $M_i$ or only between those in
the mass bin corresponding to $M_i$ itself.  As comparisons only within
a particular bin are both easier to calculate and understand physically,
however, it is this quantity that we plot in Figure \ref{fig:eta}.

Here we see that the numerical and analytical expressions agree at
lengths scales $\gtrsim R^{M_i}$, but diverge at smaller distances.
As the simulation simply excludes all points closer than $R_{M_i}$
there is a sharp fall of in $\xi_{R_{\rm M_i}}(r)$ below this
distance.  The analytical expression, however, does not exclude pairs
of objects that are contained within a larger virialized object (as would
be excluded in an excursion set calculation) and increases
dramatically at small $r$.
Thus these quantities are derived in a manner such that they must
be discrepant at small distances,  and we can have some confidence that
the overall distribution of objects in our simulations is reasonable.

While the spherical collapse model is a logical testing
ground for our code and can be simply compared to analytical predictions,
a more accurate simulation results by applying the elliptical collapse
criteria for halo formation, Eq.\ (\ref{eq:ellcriteria}).  This allows not
only for a more accurate identification of collapsed peaks on an
object-by-object basis, but delays the collapse
of the more common peaks, and thus the minimum accurate redshift at each mass
scale is decreased  to $z_{\rm min}(M_i)$ such that 
$\sigma(M_i) D(z_{\rm min}(M_i))/D_0 = 1.3 \delta_{\rm ec}$. 

In this model, the Press Schechter relation, Eq.\ (\ref{eq:press}), is modified to 
\ba
\frac{d n_{\rm ec}(M,z)}{dM} = & - A \left[1 + \left(\frac{\sigma(M)D(z)}
{\delta_{\rm sc} D_0} \right)^{2q} \right]
\sqrt{\frac{2}{\pi}} \frac{\rho(z)}{M}
\qquad \qquad \nonumber \\
 &
\frac{\delta_{\rm sc}(z)}{\sigma(M)^2 D(z)} 
\exp 
\left(
-\frac{\delta_{\rm sc}^2(z) D^2_0}{2 \sigma^2(M) D(z)^2}
\right) \frac{d \sigma(M)}{dM},
\ea
where $A = 0.322$ and $q = 0.3.$  Note that this function is essentially
equivalent to the spherical collapse predictions for rare, high-mass peaks,
and thus the $\sigma_8$ normalization from the number density of galaxy 
clusters is unchanged in this model.  This expression has been checked
against the number density of objects in our simulation and again is in
good agreement.

\subsection{Initial Gas Infall}

While our simple nonlinear collapse model captures the overall history
of the dark matter distribution, the evolution of the gas is more
involved.  While gas traces dark matter at early times, the formation
of bound objects and subsequent star formation requires that the gas
be able to cool to $T \sim 0$.

If gas collapses and virializes along with a dark matter perturbation,
and we assume an isothermal distribution, then it will be heated to
a temperature of (Eke, Cole, \& Frenk 1996)
\be
T_{\rm vir} = \frac{ 90 {\rm K}}{\beta} 
\left(\frac{6.8}{5X+3}\right)
M_6^{2/3}(1+z_c)
\left(\frac{\Omega_0 \Delta_c(z_c)}{
\Omega(z_c) 18 \pi^2}\right)^{1/3},
\ee
where $\beta$ is the ratio of specific
galaxy kinetic energy to specific gas thermal energy, $X$ is the hydrogen
mass fraction which we take to be $0.76$, $M_6 \equiv M/(10^6 M_\odot/h)$,
and $\Delta_c$ is the ratio of the mean halo density to the critical 
density at the redshift of collapse, a constant ($18 \pi^2$) for the
$\Omega=1$ case and a otherwise a weak function of $z$ fitted in
Kitayama \& Suto (1996).  Navarro, Frenk, \& White (1995) have shown that
this relation with $\beta = 1.07 \pm 0.05$
is an accurate approximation to the results of 
N-body/hydrodynamic simulations in the case of an $\Omega = 1$ universe.  
Here we follow Eke, Cole \& Frenk (1996) and fix $\beta = 1$ for all 
cosmological models.

Once the gas has fallen into a potential well, we adopt a simple model 
to calculate the time scale at which the gas will cool and form stars.
Neglecting the gravitational energy associated with further
gas infall we have
\be
n_e^2(r) \Lambda(T(r),Z(r)) = \frac{3}{2} n(r) k_B \frac{dT(r)}{dt},
\label{eq:cool}
\ee
where $n_e(r)$ is the number density of electrons at a radius $r$,
$n_{\rm tot}(r)$ is the total number density, $T(r)$ is the
temperature of the gas, $k_B$ is the Boltzmann constant, and $\Lambda$
is the radiative cooling function, which is strongly dependent on the
metallicity $Z(r)$ of the gas.  Here we take $\Lambda$ as tabulated by
Sutherland and Dopita (1993) for equilibrium configurations, although
strictly speaking these estimates are not exact due to the difference
in abundance ratios for alpha elements in SNeII outflows as
compared to solar proportions.  Considering only helium and hydrogen
the mean molecular weight is $\mu m_p = \rho_g/n = m_p 4/(8 - 5 Y)$ and
the ratio of the electron and total number density are related by
$\eta \equiv \frac{n_e}{n_{\rm tot}} = \frac{ 4(Y^{-1} -1) +
2}{8(Y^{-1}-1)+3} $, where $Y$ is the helium fraction by mass, here
taken to be $0.25.$

For a general configuration of gas, Eq.\ (\ref{eq:cool}) is difficult
to solve, and we therefore adopt a simple model for each collapsing halo
as an isothermal sphere ($\rho(r) \propto r^{-2}$) with constant metallicity.
We then follow the simple heuristic model of 
White \& Frenk (1991) in which all the gas within some ``cooling radius''
$r_{\rm cool}$ cools instantaneously and all the gas outside this
radius stays at the virial temperature of the halo, with $r_{\rm cool}$
moving outward with time.  For an isothermal sphere
this gives
\ba
\frac{dM_{\rm cool}}{dt} =&  4 \pi \rho_g(r_{\rm cool}) r_{\rm cool}^2
			\frac{dr_{\rm cool}}{dt} 
	\qquad \qquad \qquad \qquad \nonumber \\
=  & 12 f_{\rm hot}^{3/2} \left(\frac{T_{\rm vir}}{\rm K} \right)
\left( \frac{\Lambda(T_{\rm vir},Z)}{10^{-23} 
	{\rm ergs \, s}^{-1}\,{\rm cm}^3} 
       \frac{\rm yr}{t} \right)^{1/2} M_\odot {\rm yr}^{-1},
\ea
where $f_{\rm hot}$ is the fraction of the halo mass in  the form
of hot gas.  While derived from a simple model, this expression 
is in  good agreement with similarity solutions for cooling flows
given by Bertschinger (1989)  
(which are $\sim 28\%$ smaller than this expression), and the one-dimensional
simulations of Forcada-Mir\'o \& White (1996) (which are $\sim 15\%$ smaller).
We therefore follow Somerville (1997)
in multiplying the right hand side of this
equation by an overall factor $f_{\rm 0}$ which we take to be $0.8$.
Finally, we approximate $f_{\rm hot} \simeq M_{\rm baryon}/M_{\rm halo} 
= \Omega_b/\Omega_0$ 
at all times, and take the time at which $M_{\rm cool} = M_{\rm baryon}$
to be time at which a galaxy is formed.  

\begin{figure}
\centerline{ 
\psfig{file=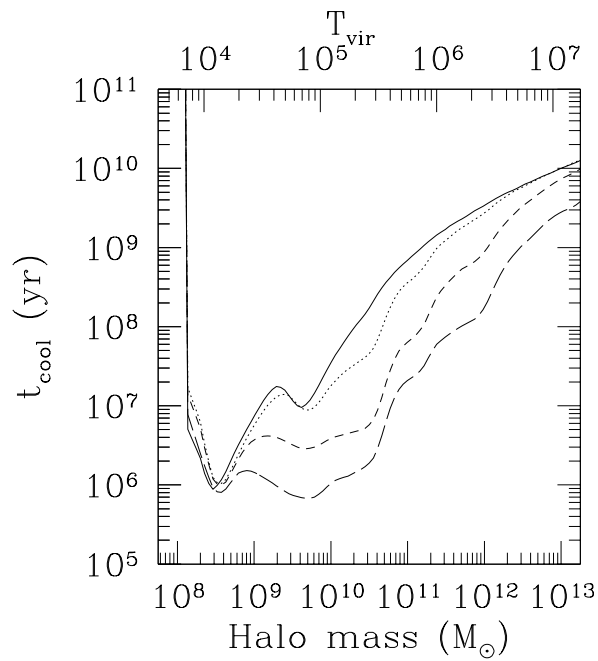,width=3.5in}}
\caption{
Cooling times adopted in our calculations  as a function of
halo mass and virial temperature at a formation redshift of 
$z = 1$, in a flat cosmology.  The solid line is for primordial
abundances ([Fe/H] = -3), the dotted line corresponds to
[Fe/H] = -2, the short-dashed line to [Fe/H] = -1, and the long-dashed 
line to solar abundances.}
\label{fig:coolz1}
\end{figure}

We show this formation time as a function of mass for four different
metallicities at $z=1$ in Figure \ref{fig:coolz1}.  From this diagram
we see that while cooling times are negligible for dwarf galaxies,
cooling times for primordial-abundance halos with masses above $\sim
10^{12} M_\odot$ are sufficiently long to prevent galaxy formation at
low redshifts.  This is dependent on metallicity, as we saw in Figure
\ref{fig:cc2}. In our simulation we therefore neglect cooling times
for halos with masses $\leq 10^{10} M_\odot$, while using the
appropriate pre-enriched cooling times for more massive objects.

Note also the sharp rise in cooling times at temperatures below
$\approx 10,000$ K, due to the lack of atomic transitions with energies below
a few eV.  Cooling at lower temperatures can only occur through
molecular transitions, and thus the collapse of $T_{\rm vir} \lesssim
10^4$ primordial gas clouds depends strongly on the existence of
molecular hydrogen.  A detailed investigation of the production and
dissociation of ${\rm H}_2$ in the early universe has been conducted
by Haiman, Rees, \& Loeb (1997).  As ${\rm H}_2$ is easily
photo-dissociated by 11.2-13.6 eV photons, to which the universe is
otherwise transparent even before reionization, they find that a UV
flux of $\lesssim 10^{-22} \,{\rm erg}\, {\rm cm}^{-2} {\rm s}^{-1}
{\rm Hz}^{-1} {\rm sr}^{-1}$, is capable of dissociating all ${\rm
H}_2$ in collapsing halos (see also Ciardi et al.\ 1999).  As this is
more than two orders of magnitudes smaller that the reionizing flux,
we assume in our simulations that a small population of early stars
quickly depleted the primordial gas of molecular hydrogen.  Thus we
exclude the formation of all galaxies with virial temperatures below
$10,000$ K, which sets the lower halo mass value to $\approx 5 \times
10^7 {\rm M_\odot} h^{-1}$, justifying the choice of spatial resolution 
for our simulations.


\subsection{Heating and Enrichment of the IGM}

Once a dwarf galaxy has been formed, we assume that a galactic outflow
is immediately generated due to SNeII from the short-lived massive
stellar population.  In order to get a rough estimate of the mass of
metals injected into the ICM, the radius over which they are mixed
into the surrounding gas, and the heating of the ICM, we model them
according to the simple analytical model developed in Tegmark, Silk,
\& Evrard (1993, hereafter TSE).  Here the authors treat the outflow
as an expanding shell that sweeps up most of the baryonic IGM and
loses only a small fraction $f_m = 0.1$ to the interior, such that the
mass of the shell is 
\be
m(t) \equiv \frac{4 \pi}{3} R(t)^3 \rho_b (1-f_m) 
\label{eq:moft}
\ee
where $R(t)$ is the radius of the outflow.
Acceleration of the shell is due to the internal pressure
and deceleration from gravitational breaking, both estimated in the
thin shell approximation (Ostriker \& McKee 1988).  With these
approximations $\ddot{R}$ becomes
\be
\ddot{R} = \frac{8 \pi p G}{\Omega_b H^2 R} 
	 - \frac{3}{R}(\dot{R} - HR)^2
	 - \Omega_0  \frac{H^2 R}{2} 
	 - \frac{GM}{R^2}.
\ee
Note that the gravitational term here includes a correction for the 
presence of a halo of mass $M$ out of which the outflow expands.
The pressure of the hot gas within the shell is given by 
$p = \frac{2 E_T}{3 V}$ where
$E_T$ is the thermal energy of the interior, and energy conservation
gives $\frac{dE_T}{dt} = L - p \frac{dV}{dt}$, where $L$ is the
luminosity, incorporating all heating and cooling of the interior
plasma.

The evolution of the blast wave is completely determined by the
assumed luminosity history of the blast wave.  Following
TSE we consider five
contributions to this luminosity: the cooling by Compton drag against
the microwave background ($L_{\rm comp}$), the cooling due to
bremsstrahlung and other two body interactions ($L_{\rm ne^2}$) as in
Eq.\ (\ref{eq:cool}), energy injection from supernovae ($L_{\rm sn}$), 
cooling by ionization of the IGM ($L_{\rm ion}$), and heating from
collisions between the shell and the IGM ($L_{\rm diss}$); such 
that,
\be
L = - L_{\rm comp} -L_{\rm ne^2} 
	+ L_{\rm sn} - L_{\rm ion} + L_{\rm diss}.
\ee
The first of these we fix according to Eq.\ (4) of TSE,
while $L_{\rm ne^2}$ is negligible in this regime.  The other
parameters require some calibration to the observed
properties of outflows from dwarf ellipticals and our overall model
of the IGM at the time of dwarf formation. Let us consider these
individually.

Following TSE we assume that the energy in the blast wave is
proportional to the number of SNeII and winds for O and B stars
and that this in turn is
proportional to the baryonic mass which has collapsed.  Energy injection
occurs over a relatively short time during which the dwarf
galaxy is undergoing a starburst phase.  We assume that a fraction of
$\epsilon_{\rm SF}$ of the baryons in each galaxy
collapse to form stars, that
one supernova is formed for every 100 $M_\odot$ of 
stars in the halo, and that each supernova provides a typical energy 
output of $10^{51}$ ergs, with an equal contribution coming from 
winds from massive stars.
If a fraction $\epsilon_{\rm wind}$ of this input
goes into the galactic outflow then the total energy from the
supernovae is $E = \epsilon_{\rm wind} \epsilon_{\rm sf} 2.0 \times
10^{55} \frac{\Omega_b}{\Omega_0} M_6.$ 
Assuming that this energy is
released at a constant rate during a period of  $t_{\rm
burn} = 1.0 \times 10^8$ years, we find
\be
L_{\rm sn} 
= {\epsilon_{\rm wind} \epsilon_{\rm sf}} 1.6 \times 10^6 L_\odot/h M_6 \times
\frac{\Omega_b}{\Omega_0}
\label{eq:lstar}
\ee
where the factor of $h$ is a result of the units we have chosen for
$M_6$.  Dekel and Silk (1986)  have proposed that the observed differences 
in surface brightness and metallicity   between the observed
classes of diffuse dwarfs and normal galaxies can be understood in
terms of outflows occurring in halos with virial velocities below
critical value on of the order of 100 km/s, which corresponds
to approximately $2.5 \times 10^9 \epsilon_{\rm SF} (1+z)^{3/2} 
{\Omega_b}/{\Omega_0}$ supernovae.  Here we take a conservative 
approach and exclude all bursts greater than $2 \times 10^7 
\epsilon_{\rm SF} {\Omega_b}/{\Omega_0},$ which is typically a
few hundred thousand supernovae and corresponds to a maximum luminosity of
\be
L_{\rm sn, max} = {\epsilon_{\rm wind} \epsilon_{\rm sf}} 3.2 \times 10^9 L_\odot \times \frac{\Omega_b}{\Omega_0}
\ee 
achieved by objects above $2 \times 10^9 M_\odot.$
 
Another potential source of drag on the expanding shell is energy
losses to ionizing the IGM, given by $L_{\rm ion} = f_{\rm neutral} \,
f_{\rm m} \, n_b \, E_o \, 4 \pi R^2 [ \dot{R} - H R],$ where $n_b$ is
the number density of baryons, $E_0 \sim 13.6$ eV, and $f_{\rm ion}$
is the ionization fraction of the IGM into which the shell is
expanding.  As they were trying to develop a model for IGM
reionization by winds, TSE assume $f_{\rm neutral} = 1$ at all times.
Here we choose for our fiducial model the case in which blast waves
expand into a medium that has completely been pre-ionized ($f_{\rm
neutral} = 0$) possibly by ionizing photons from active galactic
nuclei, or from ionization fronts from the massive stars that precede
the winds.

Finally $L_{\rm diss}$ accounts for the heating of the interior plasma
due to collisions between the shell and the IGM.  Taking this to be some
fixed fraction $f_{\rm diss}$ of the kinetic energy lost by the shell gives
\be
L_{\rm diss} = f_{\rm diss} \frac{3 m}{2 R} (\dot{R} - H R )^3.
\label{eq:diss}
\ee
In TSE the authors consider the extreme cases in which $f_{\rm diss} = 0$ 
and $f_{\rm diss} = 1$,
showing that their results are relatively insensitive to this parameter.  
Here we simply fix $f_{\rm diss} = 0.5$ throughout.

\begin{figure}
\centerline{ 
\psfig{file=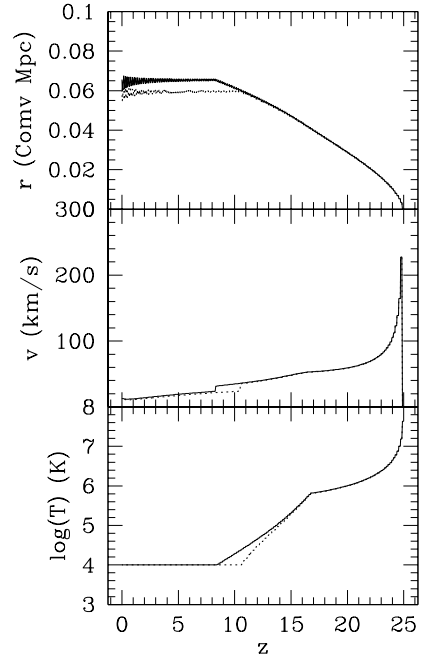,width=3.8in}}
\caption{Outflow from a $1.1 \times 10^8 M_\odot$ galaxy with 
$\epsilon_{\rm wind} = 0.1$.  The solid and dotted curves
show the expansion into an ionized and neutral medium respectively.
Note that the neutral fraction of the IGM is largely unimportant in
determining the evolution of the outflow.}
\label{fig:wind}
\end{figure}

With these approximations we can use the solution given in TSE 
to calculate the full expansion history of the galactic outflow,
with their $L_*$ and $E_*$ appropriately modified
to account for the gravitational potential of the collapsed halo
as given by our Eq.\ (\ref{eq:lstar}).  We step forward in time (backwards
in redshift) in increments of $\Delta z = 0.01$ starting at a maximum
redshift of $z_{\rm max} = 25$, using a fourth-order Runge-Kutta technique.
Figure \ref{fig:wind} shows the
first galaxy to undergo an outflow, a $1.1 \times 10^8 M_\odot$ dwarf 
expanding into an ionized medium at $z=25$ in one such simulation.


As our simulations make no attempt to include ionization and do not
account for the presence of ionizing radiation from stars within the
dwarf galaxy or an external source of UV radiation, we expect our
solution to only be accurate while the temperature of the material is
greater than 10, 000 K ($k_b T = 13.6$ eV).  After this point, we
assume as a reasonable approximation that the halo simply expands with
the Hubble flow and remains at a fixed temperature of 10,000 K.

In this figure, we also include for comparison the results of one such
blast wave expanding into a neutral medium.  This closely traces the
ionized solution up until the temperature of the outflow drops to
$\sim 10,000$ K.

Finally, we model the metallicity of the expanding halo by assuming
that each supernova typically ejects 2$M_\odot$ of metals into the 
intergalactic medium.  This number is consistent with the average stellar 
yields in SNeII simulations as compiled in Nagataki \& Sato (1997),
although there are significant theoretical uncertainties between
the various simulations.  The total mass of metal injected into the
interstellar medium of the star forming galaxies is then
\be
M_{6,z} = 0.02 \epsilon_{\rm sf} (\Omega_b/\Omega_0) M_6.
\ee

Our model has two important free parameters, 
$\epsilon_{\rm sf}$ and $\epsilon_{\rm wind}$.
While the amount of energy in the winds is directly dependent on
the product of these two parameters, their ratio is somewhat 
more unconstrained. Our approach here then
will be to fix the star formation efficiency at a somewhat standard
value of  $\epsilon_{\rm sf} = 0.1$ and allow the fraction of 
the SN energy channeled into the wind to vary between 5 and 20 percent
to quantify model uncertainties.

With this simple model we can construct from ${\cal X}^{M_i}$ and
${\cal Z}^{M_i}_{\rm c}$ the full histories of the blast waves expanding
into the intergalactic medium: the center of each expanding bubble
(${\cal X}^{\rm blast} \equiv \{{\bf x}_1^{\rm blast},
{\bf x}_2^{\rm blast},...  \}$), its comoving radius as a function of redshift
(${\cal R}^{\rm blast}(z) \equiv \{R_{1}^{\rm blast}(z),
 R_2^{\rm blast}(z),... \}$), temperature as a function of redshift
(${\cal T}^{\rm blast}(z) \equiv 
\{T_{1}^{\rm blast}(z), T_2^{\rm blast}(z),... \}$), and mass of ejected
metals
(${\cal M}_Z^{\rm blast} \equiv 
\{M_{Z,1}^{\rm blast}, M_{Z,2}^{\rm blast},... \})$.
While ${\cal R}^{\rm blast}(z)$ and ${\cal T}^{\rm blast}(z)$ are
updated in redshifts intervals of $\Delta z =0.01$, the history is
only stored in intervals of $\Delta z = 0.05$ to save memory.  Note
that the jagged appearance at late times of the comoving size of the
outflow in Figure \ref{fig:wind} is caused by this discrepancy in
times steps, rather than by a numerical instability.  Note also that
as outflows only occur in objects with masses $\lesssim 10^{10}$ for
which cooling times are negligible, we can compute the entire outflow
history of our Monte Carlo volume without considering the impact of
metal enrichment on the gas cooling times.  Suppression of galaxy
formation due to outflows from nearby objects, however, must be
considered for outflow-scale second generation objects.

\subsection{Formation of `Second Generation' Objects}

As we saw in \S2, galactic outflows affect subsequent galaxy formation
both by shocking the IGM as well as enriching it with metals.  Shocks
can suppress the formation of nearby galaxies by two main mechanisms,
heating the halo gas or striping it from the dark matter.  In the
first case the gas associated with a collapsing halo is heated to
above the viral temperature of the halo but remains bound to the
collapsing dark matter.  The thermal pressure of the gas then
overcomes the dark matter potential and the gas expands out of the
halo, preventing galaxy formation.  In the second case suppression is
caused by stripping of the baryons by a shock from a nearby source.
The momentum of the moving shock is sufficient to carry
with it the gas associated with the halo of an unvirialized perturbation, 
thus emptying the dark matter halo of its gas and  preventing a 
galaxy from forming.

In a companion letter (Scannapieco, Ferrara, \& Broadhurst 2000), 
we evaluate both these scenarios of halo suppression and find
the stripping mechanism to be dominant, because
most shock-heated clouds can radiatively cool 
within a sound crossing time.  As in the case of outflow
generation, momentum transfer between the outflows and the
collapsing halos is a complicated process, dependent on the
overall density profile of the collapsing halo and the possibility
of smaller collapsed objects within it.  Again, our approach in
this exploratory study will be to adopt simple criteria to
estimate on average when the momentum in a shell is sufficient to
remove material from a collapsing halo.

We allow for baryonic stripping to occur only if a halo has not yet
virialized, and assume that stripping occurs in all cases in which a 
shock moves through the center of the pre-virialized halo with sufficient 
momentum to accelerate the gas to the escape velocity.
Thus we exclude all objects for which
\be
	f \, M_s \, v_s \geq M_c \, v_e,	
\label{eq:strip}
\ee
where $f = \ell^2/4r_s^2$ is the solid angle of the shell that is 
subtended by the collapsing halo, $\ell$ is the radius of the
collapsing region when the shock moves through its center,   
$M_s = \frac{4 \pi}{3} \Omega_b \Omega(z)/\Omega_0 \rho_c r_s^3$ is 
the mass of the material swept up by the shock and $M_c$ is the baryonic 
mass of the collapsing halo.  
We can estimate the comoving radius of the collapsing halo as 
\be
\ell = R_{M_j} (1+\delta_{\rm NL})^{-1/3},
\ee
where $\delta_{\rm NL}$
is the nonlinear overdensity of the region, which is given in the 
spherical collapse model as
\be
1 + \delta_{\rm NL} = \frac{9}{2} 
\frac{(\theta - \sin \theta)^2}{(1 - \cos \theta)^3},
\ee
where the collapse parameter $\theta$ is given by 
$(\theta - \sin \theta)^{2/3} \pi^{-2/3} = D(z_{\rm cross})/D(z_c)$ where
$z_{\rm cross}$ is the redshift at which the shock moves through
the center of the halo.  Note that in the elliptical collapse
case, $z_c$ is computed according to Eq.\ (\ref{eq:ellcriteria}),
slowing down the condensation of the cloud somewhat.
Note also that we treat halos with subcondensations
that have already virialized in the same way.
In this case the shock would lose energy impinging on
the condensed sub-objects but would also have less uncondensed
gas to accelerate to escape velocity.  We assume for our
purposes here that these effects roughly compensate, 
although more detailed hydrodynamical studies would help
to refine this comparison.

Finally, we calculate the metallicity of each collapsing sphere by volume
averaging the contributions of each of the expanding blast waves that
pass within the collapse radius $r_{\rm col}$. Each halo is assigned a
collapse mass in metals $M_{Z,k}^{M_i}$, which is taken to be zero
initialy and modified by each blast front that passes within the
collapse radius.  For each such occurrence, the mass in metals is
updated to
\be
M_{Z,k}^{M_i} \longrightarrow M_{Z,k}^{M_i}  + 
\frac{V_{\rm overlap}}{\frac{4 \pi}{3} R_{M_i}^3}
M^{\rm blast}_Z 
\ee
where $V_{\rm overlap}$ is the volume of intersection of the two
spheres.  By dividing this mass by the total baryonic mass of the
galaxy we can compute the initial metallicity of the object.  This
value is then used to compute the collapse time of the object by using
the appropriate cooling function, $\Lambda(T,Z)$.

\section{Results}

In this section we summarize the results of our simulations, both
in relation to the suppression of low-mass galaxy formation by
baryonic stripping of pre-virialized halos, 
as well as the pre-enrichment of high mass galaxies 
by metals carried in galactic outflows.  We show that the tendency for
small galaxies to disrupt the formation of their neighbors helps to
explain the small number of Milky Way satellites relative to Cold Dark 
Matter model predictions,  and leads to a lack of small-scale galaxy 
mergers, suggesting a bell-shaped luminosity function for elliptical
galaxies.  Pre-enrichment helps to explain the high G-dwarf metallicities
found in our own and other  
galaxies, but the shorter cooling times of metal enriched
clouds seems to have little effect on the overall number of large galaxies 
formed.  We examine the effect of varying model parameters on these
features and discuss the compatibility between the star formations rates
and IGM heating in our models and constraints from optical observations
and measurements of thermal distortions in the cosmic microwave background.

\subsection{Shocks and Low-Mass Galaxy Formation}

\begin{figure}
\centerline{ 
\psfig{file=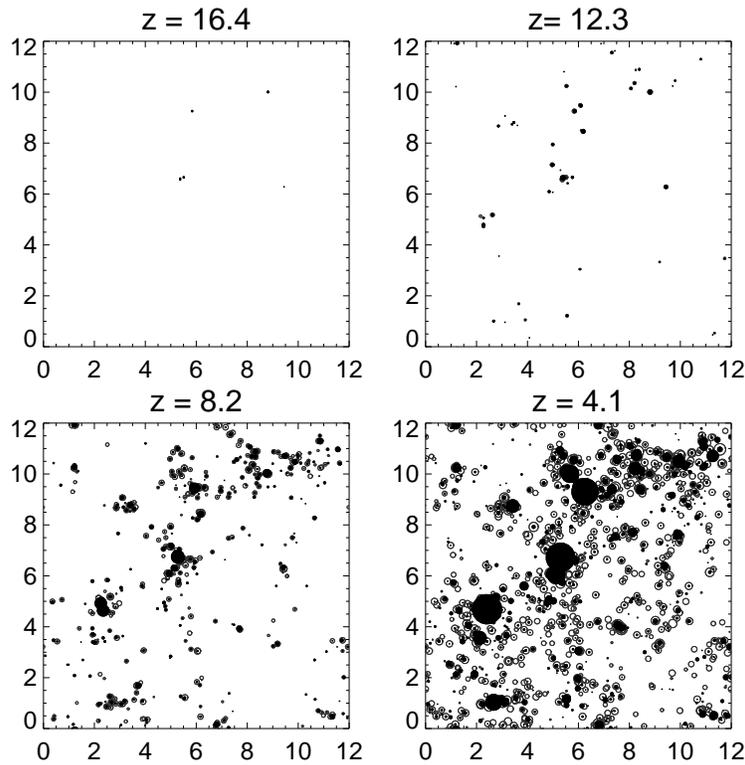,width=4.0in}}
\caption{
A slice though the simulation volume at four different redshifts.  The filled circles  show the radii of halos that have collapsed,
while the open circles show the size  of the galactic
outflows that are in progress.
Outflowing regions are highly correlated with regions in which 
new objects are collapsing at all redshifts, 
illustrating the necessity of a three-dimensional simulation.
}
\label{fig:halos}
\end{figure}

Having constructed a distribution of halos 
with collapse times and spatial
orientation in good agreement with analytical expressions, we now
consider the impact of outflows and enrichment on galaxy formation
within this distribution.  We first restrict our attention to the
$\Lambda$CDM model of Figure 1, as the currently favored cosmological
model, and consider the effects of varying the cosmology and other
model parameters in \S4.4.  In Figure \ref{fig:halos} we show the
collapse and formation of outflows in our fiducial model with
$\epsilon_{\rm wind} = 0.1$.
Here we see that regions effected by galactic outflows
are highly correlated with regions in which new objects are
collapsing, illustrating the necessity of an inhomogeneous approach.

\begin{figure} 
\centerline{ 
\psfig{file=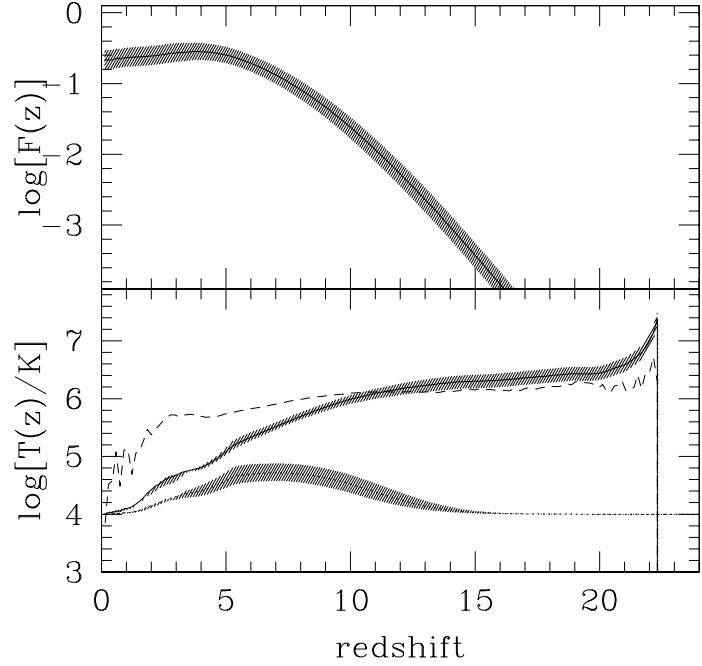,width=3.8in}} 
\caption{
{\em Top}: Volume filling factor, ${\rm F}(z)$ as a function of
redshift.  {\em Bottom:} Mean temperature within the outflowing shells
(solid line) and the simulation overall (dotted line).  The RMS
fluctuations in the temperature of outflows is given by the
dashed line.  All lines correspond to the fiducial $\epsilon_{\rm
wind} =0.1$ model, while the grey regions are bounded from above by
the high-energy outflow 
case ($\epsilon_{\rm wind} = 0.2$) and from below by the
low-energy outflow case ($\epsilon_{\rm wind} = 0.05$).}
\label{fig:por} \end{figure}

In Figure \ref{fig:por} we show the volume fraction in the expanding
shells as a function of $z$.
This quantity is not estimated on a cell-by-cell basis, but rather
approximated by its value in the case in which the positions of the
bubbles are uncorrelated,
\be
{\rm F}(z) \equiv 1 - \exp \left( - \frac{ \frac{4 \pi}{3} 
\sum_{i=1}^N R_i^{\rm blast}(z)^3}
{(12 {\rm Mpc}/h)^3} \right).
\label{eq:fofz}
\ee
Note that this quantity overestimates the volume within outflows
as the correlations between outflow positions can be significant, and
are even greater in a full N-body approach than in our simple Monte
Carlo.

Also shown on this plot is the volume-averaged temperature within
the shells and the overall volume-averaged temperature of the simulation.  
Here we see that at redshifts much greater than $\sim$ 5, the areas near
dwarf starburst galaxies were significantly hotter than the IGM as a whole.
This discrepancy is somewhat reduced at later times, when the outflows
begin to fill a significant fraction of the total volume of the universe.  
Even at these late times, however, the universe remains inhomogeneous, 
with new starbursting dwarf galaxies forming within regions only mildly 
impacted by the earliest outflowing objects.  Thus the root mean
squared (RMS) fluctuation in the temperature remains significant
at redshifts $\gtrsim$ 1.


This high degree of inhomogeneous shocking naturally has a large impact
on galaxy formation at masses below $\sim 10^{10} M_\odot$.  In Figure
\ref{fig:cpt} we plot the number of collapsed halos as a function of
redshift, along with the number of unsuppressed objects 
as described in \S3.4.  While the earliest dwarf
galaxies form along with the collapsing halos, by $z \approx 12$
outflows become important. As $M \gtrsim 10^9 M_\odot$ halos collapse
after this redshift, galaxy formation at this scale is highly suppressed,
leading to a ``mass desert'' between the major population of polluting 
dwarf galaxies and the larger
galaxies that are seen today.  It is also clear from these figures
that galaxy formation is likely to have occurred in two stages, with
the dwarf outflow population forming mostly at the highest redshifts,
and larger galaxies forming only later when halos of sufficient mass
began to collapse.  


Also in this figure, we plot the total
number density of collapsed halos with masses between $1.1\times
10^8 M_\odot$ and  $1.4\times 10^{11} M_\odot$, along with the total
number of galaxies in this mass range.  Notice that only $\sim 30 \%$ of
the halos are populated, consistent with the factor of $\sim 4$ suppression 
needed to reconcile the number of predicted and observed Milky-Way satellites
(Kauffmann, White, \& Guiderdoni 1993; Klypin et al.\ 1999; Moore et al.\ 1999).
Our scenario thus provides a natural mechanism for the formation of
``dark halo'' satellites around our galaxy, which may be associated
with the abundant High-Velocity Clouds as discussed by Blitz
et al.\ (1999).

\subsection{Enrichment and High-Mass Galaxy Formation}

In Figure \ref{fig:metal}, we plot the volume-averaged metallicity
within the outflows and the simulation overall, as well as the RMS
variation between the shells.  Here we see that like the temperature,
the mean metallicity within the outflows is the greatest 
at early times, and decreases to a few times 
the mean metallicity of the universe at a redshift of $\sim 5$.
The mean IGM metallicity from $z=1$ to $z=5$  ranges 
from about   $\log_{10}[Z/Z_\odot] \sim -1.8$ to
$\log_{10}[Z/Z_\odot] \sim -2.6$ to
in good agreement with observed metallicities in Ly absorption 
systems (e.g. Cowie \& Songaila 1998, McDonald et al 1999, 
	Ellison et al.\ 1999),
as well as previous studies of enrichment by dwarfs (Nath \& Trentham 1997).
The RMS variations in the metallicities within the outflows 
are even more severe than the temperature variations however, and remain
several times larger than the mean metallicity at all times.

This highly inhomogeneous distribution of metals is consistent with
observed metallicity inhomogeneities as discussed in \S 2.2 and
suggests that the mean metallicity of the gas out of which most
galaxies formed may be significantly higher than that of the
IGM. Figure \ref{fig:metal} also shows the initial metallicity of the
IGM newly formed galaxies.  Note that this
does not include the contribution from the metals in the progenitor
galaxies themselves.  Here we see that indeed the mean metallicity of
collapsed objects is an increasing function of mass which is
expected to be $\log_{10}[Z/Z_\odot] \sim -1$ for Milky-Way sized
objects, and  higher than the mean IGM metallicity at for all
mass scales.
This value is comparable to
the $\sim 0.1 Z_\odot$ pre-enrichment necessary to explain the lack of
low-metallicity G-dwarfs in the Solar neighborhood.
The overall relatively high initial metallicity
 is also consistent with observations 
that suggest the need for a ``floor'' of around $0.1Z_\odot$ in 
the metallicity distribution of stars in other massive galaxies.
Note that we make no attempt to trace subsequent star formation
and interstellar medium (ISM) enrichment, and
our values only provide a {\em lower limit} on the lowest
metallicity stars in disk galaxies that formed largely from the IGM
gas.  Note also that these metallicities are slowly increasing with
mass, suggesting an alternative explanation for
the mean metallicity luminosity relationship, which is usually
understood as ISM enrichment {\em within} galaxies and their progenitors
(see eg. Vader 1986; Ferguson \& Binggeli 1994;
Kauffmann \& Charlot 1998).

As collapsing halos in which galaxy formation is suppressed
are found closest to outflowing galaxies,
where both shock velocities and metallicities are high,
it is likely that whatever residual gas remains in these
objects would be of similar of even higher metallicity than
that of unsupressed objects.  While the calculation of these
metallicities is beyond the scope of our simulations,
it is interesting to note that some of the high velocity clouds may also 
have metallicities that actually exceed those of many dwarf galaxies
although may still be somewhat less than suprasolar values predicted
by a ``galactic-fountain'' model (Sembach et. al\ 1999; Wakker et al.\
1999). 


In Figure \ref{fig:cpcool} we show the effect of cooling on galaxy
formation in large-mass halos.  Here we see that while
halo collapse and galaxy formation are almost simultaneous at masses
$\lesssim 5 \times 10^{10} M_\odot$, long cooling times suppress all galaxy
formation on scales $\gtrsim 5 \times 10^{12} M_\odot$. The increased
initial metallicities help to accelerate galaxy formation 
in objects between these mass limits somewhat,
although the overall final number densities remain unchanged.


\subsection{Properties of Elliptical Galaxies}

As elliptical galaxies tend to be found in the most dense and enriched
regions of space, it is natural to expect that outflows would have had
the largest impact on these objects.  In order to study this
connection, we follow the conventional wisdom that ellipticals
correspond to mergers of (see e.g., Barnes 1992; Hernquist 1993)
large progenitors.

We therefore define $M_{{\rm merger},k}^{M_i}$ as the total contribution
to the mass of halo $k$ of mass $M_i$ due to mergers of unsupressed galaxies
of mass scales $M_{i-1}$ and $M_{i-2}$,
that is
\ba
M_{{\rm merger},k}^{M_i} & \equiv 
  \sum_{l}  \cases { M_{i-1} &  {\rm if}
	$ ||{\bf x}^{M_i}_k - {\bf x}^{M_{i-1}}_l || \leq R_{M_i}$  \cr
     		     0      &  \,\,\, otherwise        } \nonumber \\
&+  \sum_{l}  \cases { M_{i-2} &  if
	$ ||{\bf x}^{M_i}_k - {\bf x}^{M_{i-2}}_l || \leq R_{M_i}$  \cr
     	             0      &  \,\,\,{\rm otherwise}          }
\ea
where we consider only objects that have cooled and have not been
swept away by shocks while forming.
We then identify as ellipticals all 
collapsed objects with $M_{{\rm merger},k}^{M_i} \geq 0.5 M_i$,
in which half of the total mass comes from a merger of
large progenitors.

We fix this threshold in order to approximate the observed field
elliptical fraction of $\sim15\%$ (Baugh, Cole, \& Frenk 1996), and
ask the question of what mass scales correspond to these objects.

In Figure \ref{fig:ellip} we plot the total
space density of galaxies in our simulations, along with the
space density of galaxies identified as ellipticals.
In this figure we see that the mass distribution of ellipticals is
quite different than that of the overall galaxy population.  As the
outflows from dwarf galaxies suppress the formation of nearby dwarfs,
there are very few mergers at small masses, with virial temperatures much
smaller than that typical of outflows.  Thus while the number density
of halos continues to increase with decreasing mass, the number
density of ellipticals is the same at $1.4 \times 10^{11} M_\odot$ and
$4.4 \times 10^{10} M_\odot$, and decreases dramatically for even
lower masses.


Using the observed Faber-Jackson relation between the luminosity of
elliptical galaxies and the velocity dispersion, we can recast this as
an absolute r-band magnitude of $M_r = -18 - 10 \log_{10}
\sigma_{100}$, (Oegerle \& Hoessel 1991) where $\sigma_{100}$ is the
velocity dispersion in units of $100$ km/s.  This gives the luminosity
function of elliptical galaxies shown in the bottom of Figure
\ref{fig:ellip}.  While these results are necessarily crude, and are
bounded at the high-luminosity end by the finite size of out
simulations, they nevertheless naturally reproduce the observed
bell-shaped luminosity function of elliptical galaxies (see eg.,
Bromley et al.\ 1998). 

\subsection{Effects of Model Uncertainties}

We have only two free parameters in our simplified modeling, both
relating to the gas outflow, and all other modeling 
based on a simple spherical or elliptical collapse model,
involving the usual assumptions. Our conclusions are
most strongly dependent on the product of the two free parameters
$\epsilon_{\rm wind}$ and $\epsilon_{\rm sf}$, which together
determine the total energy being channeled into galactic outflows.  In
our fiducial model, we took $\epsilon_{\rm sf}$ = 0.1 and
$\epsilon_{\rm wind} \epsilon_{\rm sf} = .01$.  These canonical values
are motivated by the current observational work on dwarf galaxy
outflow energetics, but a large spread is naturally expected and thus
we have some freedom in changing this efficiency. We therefore examined two
extreme cases with $\epsilon_{\rm wind} = .05$ and $\epsilon_{\rm
wind} = .2$ to asses the robustness of the features described above.
These cases bound the shaded regions of the results plotted in
Figures, 7, 8, 11, 12.

\begin{figure}
\centerline{ 
\psfig{file=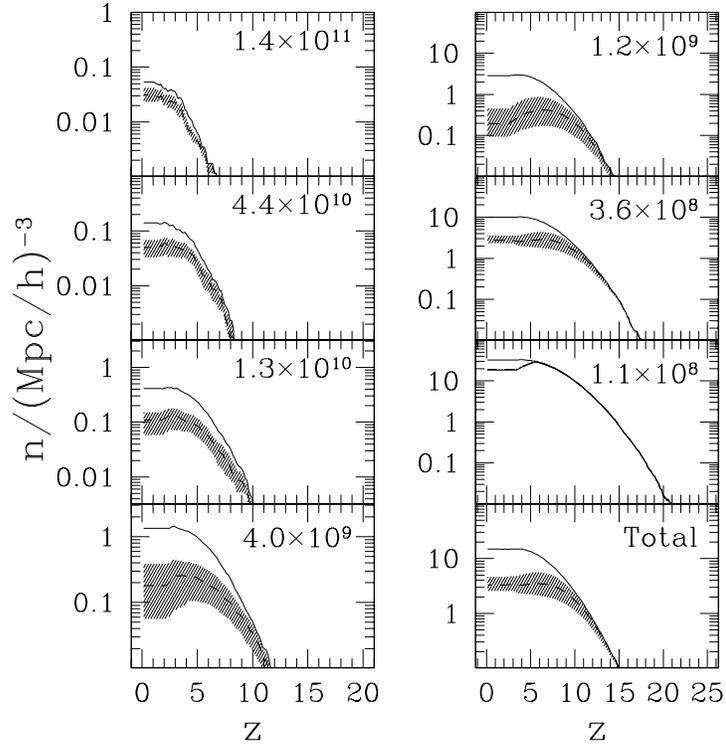,width=4.2in}}
\caption{ 
Suppression of low-mass galaxy formation by galactic
outflows.  In each panel, the upper line corresponds to the Monte
Carlo space density of halos and the dashed line corresponds to the
density of galaxies satisfying the criteria that the virial temperature
be greater than the volume-averaged temperature of the collapsing
halo, for the fiducial model.  The shaded region is bounded from above
by the low-energy outflow case and from below by the high-energy
outflow case.  The panels are labeled by mass in units of $M_\odot$.
Outflows are most important in suppressing galaxy formation in 
the mass range $10^9M_\odot - 10^{10} M_\odot$.
Note that we exclude the lowest mass scale from the total as
it undergoes suppression when its viral temperature drops below
$10,000$K at low redshifts, and the formation of these
objects is therefore much more uncertain than at larger mass scales.
}
\label{fig:cpt}
\end{figure}

\begin{figure}
\centerline{ 
\psfig{file=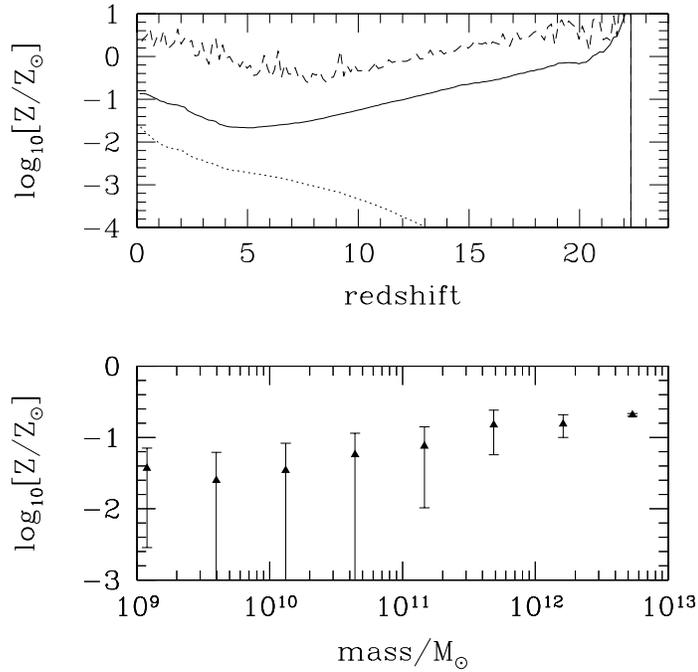,width=3.8in}}
\caption{{\em Top:} Mean metallicity within the outflowing shells (solid line) 
and the simulation overall (dotted line).  The RMS fluctuation in metallicity
within the outflows is given by the dashed line.
Note that there is a large scatter in outflow metallicities, and the 
mean metallicity within the outflows is significantly higher
the in the overall IGM average at all redshifts.
{\em Bottom:} Mean metallicity of the IGM component 
of galaxies  at the time of formation
bounded by the RMS variations in galaxy metallicity.}
\label{fig:metal}
\end{figure}

\begin{figure}
\centerline{ 
\psfig{file=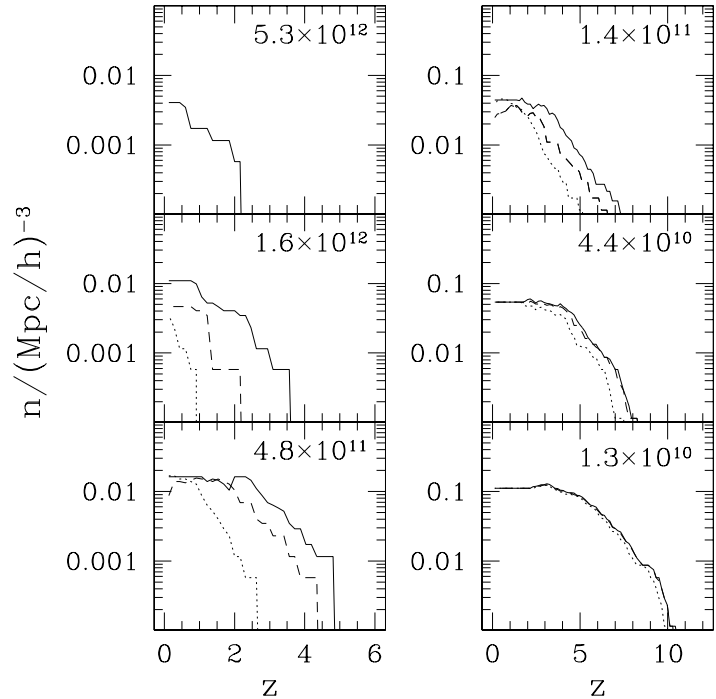,width=4.0in}}
\caption{ 
Suppression of high-mass galaxy formation due to long
cooling times.  The solid lines correspond to the Monte Carlo number
density of halos at each redshift, the dashed lines correspond to the
number density of cooled objects as calculated from primordial
abundances, and the dotted lines to the number of cooled objects with
metal enrichment. The panels are labeled by mass in units of
$M_\odot$.  Outflows accelerate galaxy formation on the $\sim 10^{12}
M_\odot$ scale, however the overall number of galaxies at $z = 0$ 
remain relatively unaffected.}
\label{fig:cpcool}
\end{figure}

\begin{figure}
\centerline{ 
\psfig{file=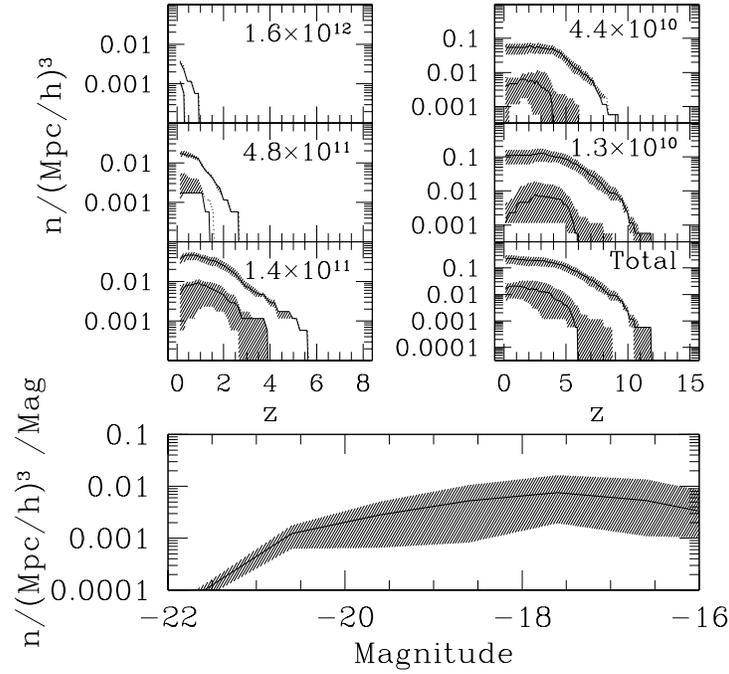,width=3.8in}}
\caption{ 
{\em Top:} Formation of elliptical galaxies.  
In each panel, the upper line
corresponds to the Monte Carlo number density of halos at each
redshift, and the lower line corresponds to the number density of
galaxies with $M_{\rm merge} \geq 0.5 M$ in the $\epsilon_{\rm wind} =
0.1$ model.  The shaded region is bounded from below by the $\epsilon_{\rm
wind} = 0.2$ case and from above by the $\epsilon_{\rm wind} = 0.05$ 
case as discussed in
\S5.4.  {\em Bottom:} Elliptical luminosity function for our
simulations.  The solid line corresponds to the total $r$-band
luminosity function for the elliptical galaxies in the fiducial model
and the shaded region is bounded by the low and high energy outflow
models.}
\label{fig:ellip}
\end{figure}

In Figure \ref{fig:por} we see the impact of model uncertainties on the
total volume fraction and temperatures of outflows in our simulations.
The impact of these differences on the suppression of dwarf galaxies
is shown in the shaded regions in Figure \ref{fig:cpt}.
Here we see that the existence of two-stage evolution with 
a ``mass desert'' in the $10^9 - 10^{10} M_\odot$ range persists 
for all values of these
parameters. The total number  of unsuppressed dwarf galaxies
ranges from $10\% - 30\%$ reinforcing the idea that outflow suppression
may be important in reproducing the small number of Milky-Way satellites.

The effect of varying $\epsilon_{\rm sf} \epsilon_{\rm wind}$ on the
distribution of elliptical galaxies is shown in Figure
\ref{fig:ellip}.  While the uncertainties in this case are somewhat
more severe than in Figure \ref{fig:cpt}, a large fall-off in the
number of objects with masses below $\sim 5 \times 10^{10} M_\odot$ is
present for all cases.

\begin{figure}
\centerline{ 
\psfig{file=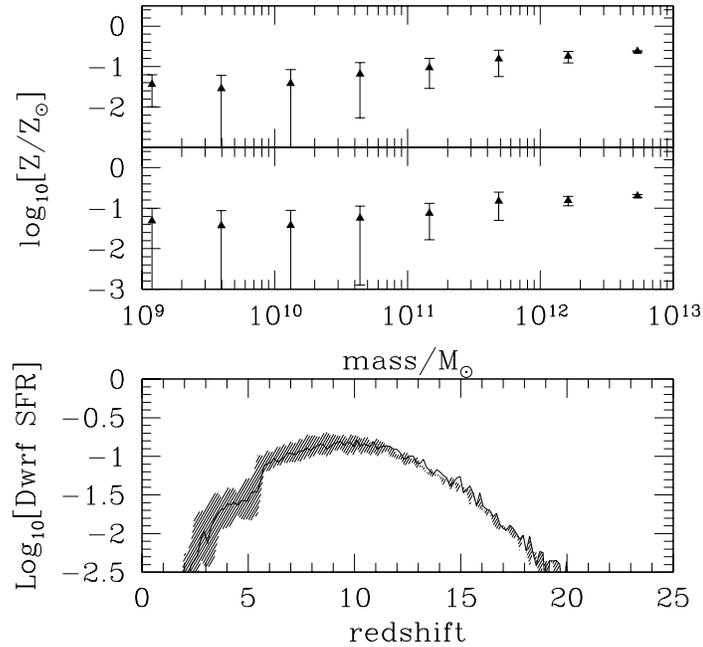,width=3.8in}}
\caption{ 
 {\em Top:} The
mean 
metallicity of galaxies bounded by the RMS variations in
galaxy metallicity.  The upper panel corresponds to the $\epsilon_{\rm
wind} = 0.05$ case and the lower panel corresponds to $\epsilon_{\rm
wind} = 0.2$.
{\em Bottom:} Star formation rate powering galactic outflows.
Again the solid line is the fiducial model and the shaded region is
bounded by the low and high energy outflow cases. }
\label{fig:met2}
\end{figure}

We show the metallicities of the IGM component of the forming galaxies
in each of these cases in the upper panel of 
Figure \ref{fig:met2}.  In both cases the
mean metallicity of collapsed objects is an increasing function of
mass which is higher than the mean IGM metallicity.  The predicted
Milky-Way scale IGM metallicity is again always
$\log_{10}[Z/Z_\odot] \approx -1$ 
and thus the G-dwarf problem is easily understood.

In the lower panel of this figure, we compute the star formation rate
in dwarf galaxies to assuage any concern that our code might be pumping
an inordinate amount of energy into the IGM.  Note that our models do
not include subsequent star-formation in disk galaxies and thus this
figure represents a lower bound to the number of stars formed at each
redshift.  Nevertheless, our values are well within observational
constraints (see eg., Adelberger \& Steidel 2000).

We can also compare our models to the COBE constraints on the
optical depth to the surface of last scatter as well as the
the distortion of the cosmic microwave background (CMB) spectrum
due to the Sunyaev-Zel'dovich effect.  We can estimate the optical
depth in our model as
\be
\tau = 	5.9 \times10^{-3} \frac{\Omega_b}{\Omega_0}
	\int_0^{z_{\rm max}} dz \frac{dx}{dz} \Omega(z) \, (1+z)^2 {\rm F}(z),
\ee
where $x$ is again the comoving distance defined such that $a_0 H_0 =
c$.
This analysis gives $\tau = 0.05$ for
the fiducial model, and varies from $\tau = 0.04$ to $\tau = 0.06$ in
the low and high energy outflow cases, which are all well within the
observational limit of $\tau \lesssim 0.5$ (Griffiths, Barbosa, \&
Liddle 1999).

The degree of CMB spectral distortions is given by the Compton-$y$
parameter which is the convolution of the optical depth with the
electron temperature along the line of sight (Zel'dovich \& Sunyaev
1969; Sunyaev \& Zel'dovich 1972).  Thus
\be
y = 1.0 \times10^{-8} \frac{\Omega_b}{\Omega_0}
	\int_0^{z_{\rm max}} dz \frac{dx}{dz} T_5(z) \Omega(z) \, (1+z)^2 {\rm F}(z),
\ee
where $T_5(z)$ is the mean temperature within the outflows as a function
of redshift in units of $10^5$K.  In this case $y$ is 
$1.6 \times 10^{-6}$ in the fiducial model, and  
varies from $1.0 \times 10^{-6}$
to $2.3 \times 10^{-6}$ for the extreme cases. 
 These values are again within
the observational constraint of $y \leq 1.5 \times 
10^{-5}$ (Fixsen et al.\ 1996).  Nevertheless, our modeling suggests
that the spectral distortions due to galactic outflows should be observable
with the next generation of cosmic microwave background experiments.

Varying the second parameter in our model, $\epsilon_{\rm sf}$,
while keeping $\epsilon_{\rm sf} \epsilon_{\rm wind}$ 
fixed acts as an overall shift
in the star formation rate and metallicities in Figures
\ref{fig:metal} and \ref{fig:met2} and can be estimated
simply by eye.  Including the ionization drag term in  Eq.\
(11) and allowing the shocks to act as the ionizing source of the
IGM has little effect on our results as suggested by Figure
\ref{fig:wind} and verified by a full simulation.
Thus we see that all the major features of the fiducial
model persist over a wide range of parameters.

We have also examined the cosmological dependence of our results
by examining a flat model
with parameters as described in \S 2.3 and $\epsilon_{\rm wind} =
0.1$.  In this case only $\sim 50 \%$ suppression of galaxies occurs,
again with the most impact on galaxies in the few times $10^9M_\odot$
range.  In this model elliptical formation falls off below the 
$10^{11} M_\odot$ scale but more gradually than in the open
case, and initial Milky-Way metallicities again about
$\log_{10}[Z/Z_\odot] = -1.0$.  Thus even in a model in which
structure forms much more quickly than suggested by observations (see
eg., Bahcall, Fan, \& Cen 1997), the major features in our model
persist.

\section{Discussion}

Our treatment of galactic outflows is intended as an exploratory
study of these processes, and makes a number of approximations
that should be made explicit.
While a Monte Carlo approach uncovers many of the issues of
inhomogeneity that can not properly be studied analytically, it fails
to capture the nonlinear structure and clustering present in
 a full N-body
treatment. At late times peaks will be grouped along sheets and
filaments, and thus clustering is likely to be more severe than in our
simulations and of a more complicated nature.

A number of uncertainties also arise from our simple treatment of IGM
enrichment and shocking.  One concern is the presence of radiative
feedback from ionization fronts around dwarf galaxies.  Our
assumption that fronts precede the formation of outflows, and thus
each shell can be treated as expanding into an ionized medium is
probably a reasonable one, while our placement of a ``temperature
floor'' at the ionization temperature of hydrogen at all times in our
simulations is stronger approximation.  Fortunately the extreme
fragility of molecular hydrogen suggests that the formation of objects
with virial temperatures below $\sim$10,000 K may have not been largely
affected by reionization.  While the inhomogeneous structure of
reionization is likely to have had a great impact on the CMB
fluctuations (Aghanim et al.\ 1996; Miralda-Escude, Haehnelt, \& Rees
2000; Scannapieco 2000), the formation of $10^7 M_\odot$ galaxies is
likely to have been halted by the dissociation of molecular hydrogen
by the first stars, long before reionization took place.

A bigger concern may be the structures of outflows and the ejection
fractions of interstellar gas and metals.  This has been studied in
detail by Mac Low \& Ferrara (1999) and Ferrara \& Tolstoy (2000), who
conclude that efficient ejection of the interstellar medium or
``blowaway'' occurs only in halos with masses $\lesssim 10^7 M_\odot$.
The issue of whether dwarfs retain a sizeable fraction of their gas,
however, is to some degree decoupled from the formation of outflows.
In halos in the mass range $10^7 M_\odot \lesssim M \lesssim 10^9
M_\odot$, a ``blowout'' occurs in which the super-bubbles around
groups of SNeII punch out of the galaxy, shocking the surrounding IGM
and efficiently ejecting metals while failing to excavate the
interstellar medium of the galaxy as a whole.  Such a scheme in which
outflows are formed primarily of the enriched IGM surrounding dwarf 
galaxies is equivalent to outflows of galactic gas for the
purposes of our simulations.  Additionally such a picture
would help to reconcile 
observations of expanding high metallicity shells around dwarf galaxies 
as described
in \S2 with observations of multiple episodes of and HI gas in dwarf
spheroidal galaxies (Smecker-Hane et al.\ 1994; Grebel 1998) some of
which even suggest that many of these objects are gas-rich, but with
extended HI envelopes (Blitz \& Robishaw 2000).

That being said, the TSE model adopted for outflow evolution in our
simulations is almost certainly oversimplified.  The ``blowout''
scenario described by Mac Low \& Ferrara (1999) ejects matter
perpendicular to the thick disk, and is thus quite
different from the spherical outflow modeled here.  Even if the winds
from dwarf galaxies can be reasonably approximated as spherical
shells at early times, such outflows of low-density heated gas
would necessarily become Rayleigh-Taylor unstable as they expanded
into the denser IGM.  Thus material is most likely to flow out in a
number of heated clumps at large distances.

Also ignored in our model are shell-shell interactions which would
contribute additional heating but slow expansion in regions
which two shocks meet.  Similarly, an additional source of IGM
enrichment is the merger mechanism described in Gnedin \& Ostriker
(1997) and Gnedin (1998), in which a significant fraction of the
interstellar medium of merging galaxies is ejected.

Finally, the criteria for halo suppression used in our simulation is
over-simplified.  While regions of space in our
simulations that meet the stripping criterion, Eq.\ (\ref{eq:strip}),
are likely to have delayed or suppressed galaxy formation, the
one-to-one correspondence assumed in our simulations is unlikely.  In
reality the formation of galaxies is likely to be a complicated
function of the number and strength of the shocks moving through the
regions, and the structure of these regions during and after halo
collapse.  This subject merits further investigation and would 
help to sharpen our conclusions.

\section{Conclusions}

While the impact of preheating and
enrichment on the observed properties of galaxy clusters has long
been recognized, the impact on the observed properties of galaxies
themselves has been little explored.  By accounting for
the observed outflows from dwarf galaxies, we have
been able to show how many of the unexplained properties of
galaxies and the IGM can be naturally understood.

Firstly, the suppression of low-mass galaxy formation by outflows
provides a natural explanation for the factor of $\sim 4$ discrepancy
between the number of observed Milky-Way satellites and predictions
from standard CDM models that do not include outflows.
Suppression also provides a natural mechanism for the formation of
``dark halos'' which may be associated with the High-Velocity clouds.

Secondly, baryonic stripping results in a bell-shaped luminosity
function of ellipticals. The lower the mass of a halo, the more likely
it is to generate an outflow that strips material from a similar mass
neighboring pre-virialized halo that would otherwise later form into a
galaxy.  This results in very few pairs of neighboring low-mass
galaxies, and hence a relative deficit of major low mass-mergers.

Finally, our models of enrichment of protogalactic gas predict a trend
of increasing metallicity with galaxy mass in good agreement with
inferences from observations.  The initial metallicity predicted for a
Milky-Way mass galaxy is $\sim 0.1 Z_\odot$ providing a natural initial
floor at the level required to solve the G-dwarf problem and the more
general lack of low metallicity stars in well studied massive
elliptical galaxies relative to ``closed-box'' models of chemical
enrichment.

These results are persistent over a wide range of model parameters and
cosmologies, and are not a result of fine-tuning parameters or
invoking additional physics. 

While galaxy outflows are already incorporated into modern studies
of galaxy formation, this is done only as an internal modification,
ignoring pre-enrichment.  Keeping track of the effect of outflows on
neighboring halos is essential in understanding the properties of
galaxy clusters, and hence it is not surprising that these effects
would play a major role in the formation of galaxies as well.
While the details await further investigation, it is clear that
any complete picture of galaxy formation must account for heating and
enrichment.

\acknowledgments

We would like to thank Rychard Bouwens, Andrew Cumming, Julianne
Dalcanton, Marc Davis, Andrea Ferrara, Ignacio Ferreras, Brenda Frye,
John Huchra, Siang Peng Oh, Alvio Renzini,
David M. Sherfesee, Joseph Silk, Jonathan C. Tan,
Robert Thacker, and Simon White for helpful comments and discussions. 
This research was supported in part by the National Science Foundation
under Grant PHY94-07194.  TJB acknowledges NASA grant AR07522.01-96A.

\newpage


\newpage

\end{document}